\documentclass[useAMS,usenatbib,usegraphicx]{mn2e}
\usepackage[T1]{fontenc}
\usepackage{aecompl}
\usepackage{times}
\usepackage{journal_names}
\usepackage{lscape}

\title[NIR Galaxies in the ZoA]{Deep NIR photometry of H\,{\Large\bf I} galaxies in the Zone of Avoidance}
\author[Williams et al.]{W. L. Williams\thanks{E-mail:
wwilliams@strw.leidenuniv.nl (WLW)}$^{1,2,3}$, R. C. Kraan-Korteweg$^{1}$ and P. A. Woudt$^{1}$\\
$^{1}$Department of Astronomy; Astrophysics, Cosmology and Gravity Centre (ACGC), University of Cape Town,\\
 Private Bag X3, Rondebosch 7701, South Africa\\
$^{2}$Leiden Observatory, University of Leiden, PO Box 9513, 2300 RA Leiden, The Netherlands\\
$^{3}$Netherlands Institute for Radio Astronomy (ASTRON), PO Box 2, 7990AA, Dwingeloo, The Netherlands
}

\newcommand{\marc}{mag~arcsec$^{-2}$~}

\newcommand{\kph}{km\,s$^{-1}$}
\newcommand{\dg}{\degr}
\newcommand{\m}{^{\rm m}}
\newcommand{\HI}{H\,{\sevensize{I}}} 
 \newcommand{\reply}[1]{{#1}}

\begin{document}

\date{Accepted 2014 June 9. Received 2014 June 2; in original form 2014 January 15}

\pagerange{\pageref{firstpage}--\pageref{lastpage}} \pubyear{2014}

\maketitle

\label{firstpage}

\begin{abstract}
\reply{Current studies of the peculiar velocity flow field in the Local Universe are limited by either the lack of detection or accurate photometry for galaxies at low Galactic latitudes. The contribution to the dynamics of the Local Group of the largely unknown mass distribution in this `Zone of Avoidance' remains controversial. We present here the results of a pilot project to obtain deep near infrared (NIR) observations of galaxies detected in the systematic Parkes deep \HI\ survey of the ZoA -- $578$ galaxies with recession velocities out to $6000$\,\kph\ were observed with the 1.4\,m InfraRed Survey Facility SIRIUS camera providing $J$, $H$ and $K_s$ imaging $~2\m$ deeper than 2MASS. After star-subtraction, the resulting isophotal magnitudes and inclinations of ZoA galaxies are of sufficient accuracy (magnitude errors under $0.1$~mag even at high extinction) to ultimately be used to determine cosmic flow fields ``in'' the ZoA via the NIR Tully-Fisher relation. We further used the observed NIR colours to assess the ratio of the true extinction to the DIRBE/IRAS extinction deep into the dust layers of the Milky Way. The derived ratio was found to be $0.87$ across the HIZOA survey region  with no significant variation with Galactic latitude or longitude. This value is in excellent agreement with the completely independently derived factor of $0.86$ by Schlafly \& Finkbeiner based on Sloan data far away from the Milky Way.} 
  
\end{abstract}


\section{Introduction}
The mass density field in the local Universe can be inferred from the measured peculiar velocities of galaxies, independent of any a priori assumption about the bias between visible and dark matter \citep{BertschingerDekel1989,Dekel+1990,Dekel1994}. The determination of the peculiar velocity field requires large and uniform galaxy samples with high-fidelity redshift-independent distance measurements.  However, peculiar velocity surveys are plagued by poor detection statistics in the Zone of Avoidance (ZoA) where the obscuring effects of dust and stars in the Milky Way prevent the identification of galaxies across $10-20$~per~cent of the sky \citep[e.g.][]{KKLahav2000,KK2005}.  This problem is usually circumvented by statistical interpolation of the mass distribution adjacent to the ZoA \citep[e.g.][]{Yahil+1991,Lynden-Bell1989,Lahav+1994}. However, some studies, such as those by \cite{Kolatt1995} and \cite{Loeb2008}, suggest that these interpolations are inadequate and require unknown mass distributions to satisfactorily explain the peculiar motion of the Local Group with respect to the Cosmic Microwave Background (CMB). \reply{On the contrary, other works do not require any missing mass distributions to  account for the local group motion \citep[e.g.][]{ErdogduLahav2009}}. Several dynamically important structures, including the Great Attractor \citep[GA;][]{Lynden-Bell1988} and Local Void \citep{TullyFisher1987} are known to reside within the ZoA, and remain poorly mapped. Indeed, it is still a contentious  issue -- 25 years after its discovery -- whether the Great Attractor is at rest with respect to the CMB
, or itself partakes in a flow towards more distant attractors like the Shapley Concentration, possibly in combination with the Horologium-Reticulum Supercluster \citep{Hudson.et.al.2004,Lavaux.et.al.2010,Lavaux.Hudson.2011}. 

Until we understand the mass distribution in the nearby Universe, {\sl including the Zone of Avoidance},  we can not properly resolve the controversial results on what fraction of the bulk flow is generated  locally (within about $\sim 100$~Mpc), and what fraction results from inhomogeneities on even larger scales \cite[e.g.][]{Hudson.et.al.2004,Watkins.et.al.2009,Feldman.et.al.2010,Macaulay.et.al.2011,Bilicki.et.al.2011,Nusser.Davis.2011,Abate.Feldman.2012,Kashlinsky.et.al.2010,Kashlinsky.et.al.2013}. A full census of the large-scale structures hidden behind the Milky Way, preferentially with a map of the underlying ZoA density field, respectively cosmic flow field, is an absolute necessity to warrant that uncertainties due to the ZoA do not perpetuate into deeper whole-sky survey results.

Blind \HI~surveys have been shown to be most effective at detecting gas rich galaxies in the most obscured parts of the ZoA. The \HI~Parkes Deep Zone of Avoidance Survey \citep{Henning2005}, conducted  on the $64$~m Parkes Radio Telescope in Australia\footnote{The Parkes telescope is part of the Australia Telescope which is funded by the Commonwealth of Australia for operation as a National Facility managed by CSIRO.}, detected over $1000$ galaxies in the southern ZoA. With an  exposure time five times longer than HIPASS, the average rms noise of the
survey was 6~mJy~beam$^{-1}$. It covered a velocity range of $-1200<v<12\,700$~\kph\ with a channel spacing of $13.2$~\kph. The survey covered the entire southern ZoA visible from Parkes: $212\degr \le l \le 36\degr$, $|b| < 5\degr$  \citep{Henning2005};  followed by an extension to the north $36\degr < l < 52\degr$ and $196\degr < l< 212\degr$, $|b| < 5\degr$
\citep[Northern Extension;][]{Donley2005}; with a later broadening of the width of the survey around the Galactic Bulge of $332\degr < l < 36\degr$, $5\degr < |b| < 10\degr$ and $352\degr<l<24\degr$, $10\degr<b<15\degr$  \citep[Galactic Bulge extension;][]{Nebthesis}. The \HI~observations reveal galaxies where both optical and NIR surveys fail  due to high extinction and high stellar density. In fact, the deep Parkes \HI\ ZoA survey region was chosen specifically to cover this otherwise impenetrable region of the sky. 

Apart from the large-scale structures delineated by galaxies detected  in \HI, their distances can be determined through the relation between linewidth and luminosity \citep[TF;][]{TF1977}, and therewith their peculiar velocities. TF distances require three observational parameters: a measure of the flux and inclination of the galaxy as well as a measure of the rotational speed of the galaxy. The latter is readily available from the \HI~linewidths\reply{, corrected for inclination obtained from the surface photometry}, while the former can be determined directly through surface photometry. Several such large-scale TF projects have been conducted to study the dynamics in the local universe, including the SFI \citep[e.g.][]{Giovanelli+1994}, expanded to the SFI++ \citep{Masters+2006,Springob+2007}, the Kinematics of the Local Universe \citep[KLUN; e.g.][]{Bottinelli+1992} and KLUN++ \citep{Theureau+2007}. 

More recently the 2MASS Tully-Fisher project \citep[2MTF][]{Masters+2008} has been launched with the goal to perform a whole-sky TF analysis based on a complete sample of 2MASS galaxies brighter than $K_{s}^o < 11\fm25$ within a volume delimited by $v < 10\,000$ \kph. The use of NIR photometry for the TF relation has several advantages over optical photometry \citep{AHM1979}:
\begin{enumerate}
 \item \label{step:ext} The extinction, both internal and Galactic, is significantly reduced in the NIR. This results in smaller uncertainties introduced by extinction corrections and more robust photometric measurements. Also, effects on the inclination due to extinction are minimised.
 \item The spectral energy distribution of galaxies peaks in the NIR, where it is dominated by the old, red stellar population. The long-lived low mass stars provide the best indicator of the total stellar mass which governs the kinematics of galaxies and is unperturbed by recent short-term star formation which may be significant at bluer wavelengths.
 \item The distribution of old stars is smoother, resulting in a smooth light profile. This makes it easier to fit elliptical isophotes resulting in cleaner photometry.
\end{enumerate}

Most importantly for our efforts in unveiling the ZoA is the low extinction in the NIR, compared to shorter wavelengths. This makes the NIR TF relation most ideal for use in the ZoA. However, the ``whole-sky'' 2MTF did exclude the innermost ZoA ($|b| < 5\degr$), as did the preceding 2MASS Redshift Survey \citep[2MRS;][]{Huchra.et.al.2012} from which the 2MTF was extracted.  The reason is twofold: it is notoriously hard to get redshifts of optically obscured galaxies, even if visible in the NIR; secondly, 2MASS is highly incomplete around the Galactic Bulge area at these low latitudes, and therefore along most of the southern ZoA which encompasses the dynamically important local large-scale structures such as the Great Attractor and the Local Void.  

The finalisation of the deep Parkes HIZOA survey has finally managed to trace  the missing large-scale structures in the southern ZoA -- although sparsely --  through the detection of the gas-rich galaxies. The remaining question is, can we also get accompanying NIR imaging for these galaxies to permit a NIR TF analysis? Only few of these galaxies have counterparts in 2MASS (about 20-30 per cent, depending on Galactic longitude), and these lack the depth and resolution to accurately measure their photometric parameters. Deep high-resolution NIR observations are, however, able to penetrate the dust and deblend foreground stars, making it possible to detect NIR counterparts for the \HI-detected spiral galaxies and measure their magnitudes and inclinations. As a pilot project,  we have therefore started a NIR follow-up survey of a substantial sample of the HIZOA galaxies across the entire southern Galactic plane, delimited by $v < 6000$~\kph. This paper presents the results of this survey. It is the first step of a larger ongoing campaign to map the large-scale structures along the full 360\degr-circle of the ZoA \citetext{Henning et al. in prep, Kraan-Korteweg et al. in prep.}, and follow up with a ZoA-optimised NIR TF analysis \citetext{Said et al. in prep.}, now that this pilot project has proved this to be a feasible approach.

This paper is organised as follows. Section \ref{sect:obs} describes the NIR observations, data reduction and image calibration. In Sect. \ref{sect:phot}, we discuss the detection of galaxies in the images, the removal of foreground stars as well as the surface photometry and determination of photometric parameters. The catalogue is analysed and discussed in Sect. \ref{sect:discuss}, including an analysis of the variation in extinction across the ZoA and a calibration of the DIRBE extinction maps.

\section{Observations and Reduction}
\label{sect:obs}
\subsection{Observations}
The imaging data were acquired with the Japanese InfraRed Survey Facility (IRSF), a $1.4$\,m Alt-Azimuth Cassegrain telescope situated at the South African Astronomical Observatory (SAAO) site in Sutherland, South Africa. The IRSF is equipped with the Simultaneous InfraRed Imager for Unbiased Surveys (SIRIUS) capable of simultaneous imaging in the three near infrared bands  $J$, $H$ and $K_s$. The camera consists of three $1024\times1024$~pixel HgCdTe (HAWAII) arrays each with a gain of $5.5~e^{-}/$ADU and a read-out-noise of $30~e^{-}$ \citep{Nagashima1999,Nagayama2003}, cooled to $80$~K. The field-of-view of is $7\farcm7 \times 7\farcm7$ and the pixel scale is $0\farcs45$~pix$^{-1}$.  The high resolution of the IRSF greatly helps in identifying and deblending foreground stars, allowing them to be removed  more accurately and improving the surface photometry of galaxies. Based on previous experience with  SIRIUS in the ZoA \citep{Nagayama2004,Nagayama2006,Skelton2009,Riad2010} each target field was observed with 25 frames of 24~s, resulting in an effective exposure time of 600~s. The resulting limiting magnitude is approximately $2$~mag deeper than 2MASS \citep{Riad2010}.  This exposure time is long enough for the detection of fainter extragalactic objects while not so long that the field is saturated with the fainter old stellar population within our Galaxy. The field-of-view of the IRSF is well matched to the positional accuracy ($\la 4\arcmin$);  of the HI sources \citep{Donley2005}; see also Fig.~9 in Sect.~4.1.

The observations for this pilot follow-up survey were started in 2006 and were continued through to 2010. Table \ref{tab:observations} lists the progress of the observations and the observers who contributed to the survey. Some fields were observed as part of the Norma Wall Survey \citep{Riad2010}. A total of $7$ weeks were allocated exclusively to this project, however a significant amount of time ($\sim60$~per~cent) was lost due to bad weather and serious problems with the detector cooling system in 2010.

\begin{table}
 \centering
 \begin{minipage}{80mm}
  \caption{Summary of IRSF/SIRIUS observations}
\label{tab:observations}
  \begin{tabular}{@{}cccc@{}}
  \hline
Year & Month(s) &  Fields &  Observer(s)\footnotemark[1] \\ 
\hline
2006 &	March $-$ June & 	7 &	NWS\\
2007 &	March $-$ May & 	50 &	NS/JT/NWS\\
2007 &	August/September & 	8 &	NWS\\
2008 &	January & 	7 &	EE\\
2008 &	March/April & 	7 &	NWS\\
2009 &	March/April & 	297 &	WW\\
2009 &	June & 	67 &	WW\\
2010 &	February & 	13 &	PK\\
2010 &	June/July & 	124 &	WW\\
\hline
\multicolumn{4}{l}{\footnotesize\textsc{Notes:}}\\
\multicolumn{4}{p{\textwidth}}{\footnotesize\footnotemark[1]Norma Wall Survey (NWS), Nebiha Shafi (NS), James Tagg (JT), Ed Elson (EE), Wendy Williams (WW), Paul Kotze (PK)}\\
\end{tabular}
\end{minipage}
\end{table}

\begin{figure*}
\includegraphics[width=\textwidth]{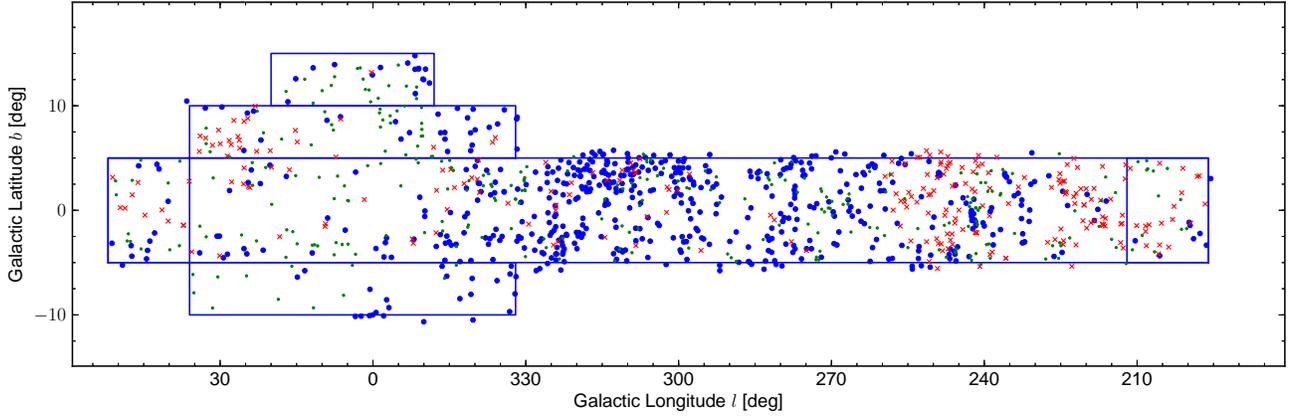}
  \caption[Near-infrared survey completeness]{Observations completed within the allocated observing runs. The blue rectangles mark the HIZOA survey region. Of the $825$ galaxies with $v\leq6000$~km\,s$^{-1}$ selected for observation $578$ ($70$~per~cent) have been observed and are plotted with large blue circles. The remaining $247$ are plotted with red crosses. The positions of the $286$ HIZOA galaxies with $v>6000$~km\,s$^{-1}$ are indicated with small green points.}
\label{fig:HIZOAobs}
\end{figure*}

\subsection{Target Selection}
\begin{figure}
\includegraphics[width=0.45\textwidth]{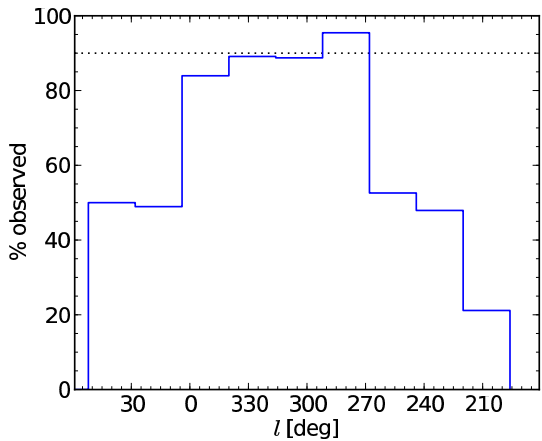}
  \caption[Survey completeness per longitude]{The percentage of galaxies observed out of the sample with $v<6000$~km\,s$^{-1}$ plotted in $20\dg$ bins of Galactic longitude. The relative incompleteness of the observations at both $l < 270\degr$ (Puppis/Hydra-Antlia filaments) and $l>0\degr$ (Local Void) is evident, while the observations around the Great Attractor region are nearly complete.}
\label{fig:HIZOAobs_l}
\end{figure}

Targets for the dedicated follow-up NIR observations were selected from a preliminary version of the HIZOA survey catalogue, containing $\sim 1100$ sources. Our targets were further selected to lie within $6000$~\kph. The reasons for this cut-off are: 
\begin{enumerate}
 \item this volume encompasses most of the large-scale structures in the nearby universe, including the Centaurus cluster, the Norma Wall/Great Attractor and the Hydra cluster \citep{Radburn-Smith+2006};
 \item at this cut-off HIZOA is still sensitive to $M_{HI}^*$ galaxies;
 \item  an $M^*_{K_s} = -24.16$~mag galaxy \citep{Kochanek2001} at $6000$~\kph~will have a apparent magnitude of $10.35$~mag. Even with extinction of $A_{K_s}=1$~mag ($A_B \approx 10$~mag), we should be able to detect galaxies up to $3.5$~mag fainter than $M^*_{K_s}$ given a limiting magnitudes of $15.0$~mag \citep[for the NWS;][]{Riad2010};
 \item the mass density \reply{within this volume} and within the ZoA may have a significant effect on the motion of the Local Group \citep{Kolatt1995,Loeb2008}; and
 \item approximately 70 per cent of the HIZOA sources lie within this volume, giving us dense enough coverage to reveal a detailed flow field in the ZoA.
\end{enumerate}
Fig.~\ref{fig:HIZOAobs} shows the spatial distribution of all HIZOA candidate galaxies; the targets selected for NIR follow-up observations, i.e. those within $6000$~\kph, are plotted with large blue points if they were observed and with red crosses if they were not observed during the allocated observing time. More distant HIZOA sources are plotted with small green points. The completeness fraction of the observations per longitude range is shown in Fig.~\ref{fig:HIZOAobs_l}. The relative incompleteness of the observations at both $l < 270\degr$  and $l>0\degr$ is largely a result of the observing time lost due to poor weather and technical problems. The observations in the Norma region are nearly complete which is excellent for studying the flow fields around the Great Attractor. Unfortunately the completeness is $\sim50$~per~cent or less around the Local Void and Puppis regions ($l < 0\degr$ and $l > 270\degr$ respectively). The impact of the low coverage for $l\la 250\degr$ is somewhat compensated for by the better detection rate in the 2MASX in this region due to the lower stellar density.

\subsection{Data Reduction and Calibration}
\label{sect:reduce}
The data were reduced using the \textsc{sirius} pipeline  \citetext{Nakajima priv. comm.}, which is a dedicated \textsc{iraf}\footnote{Image Reduction and Analysis Facility (\textsc{iraf}) is distributed by the National Optical Astronomy Observatory (NOAO), which is operated by the Association of Universities for Research in Astronomy, Inc, under cooperative agreement with the National Science Foundation} package for the automated reduction of IRSF/SIRIUS images. The high-level steps that encompass the standard NIR image reduction include:
\begin{enumerate}
  \item dark current subtraction;
  \item determination of master flat fields and flat correction;
  \item sky determination and subtraction; and
  \item frame to frame offset determination and combination.
 \end{enumerate}

Sky frames are determined from a \textit{self-sky} image produced by median combining the individual dithered frames without realigning them. Because detector artefacts remain fixed on the detector while astronomical sources move with the dither pattern, any features (e.g. stars) on scales less than the spacing between dithered frames will be removed, leaving only the sky structure. However, extended astronomical objects with scales larger than the field separation will leave a residual on the sky image.

The output images from the \textsc{sirius} pipeline are astrometrically and photometrically calibrated against the 2MASS Point Source Catalogue \citep[2MPSC;][]{Strutskie2006}. The calibration is done with a combination of \textsc{iraf} and \textsc{python} scripts developed by N.~Matsunaga\footnote{Institute of Astronomy, School of Science, University of Tokyo} and modified by \cite{Riad2010}. The astrometric calibration uses the method of Optimistic Pattern Matching \citep[OPM;][]{Tabur2007} to match stars in the 2MPSC with point sources in each image. The astrometric solution is  found using the \textsc{iraf} task \textsc{ccmap}. We use J2000 coordinates.

The photometric calibration makes use of an \textsc{iraf} script developed by \cite{Riad2010} which matches point sources within $1\farcs35$ ($3$ pixels) in each image with sources in the 2MPSC. The instrumental magnitude, $m_{\lambda}^i$, is transformed to a standard magnitude by
\[
 m_{\lambda} = ZP^i_{\lambda} + m_{\lambda}^i,
\]
where $ZP_{\lambda}$ is the magnitude zero-point. The uncertainty in the zero-point of the standard star calibration was typically 0.06 mag in all three bands. The SIRIUS filters \citep[][]{Tokunaga2002} are slightly different from the 2MASS filters so a transformation is required between the two systems. This is applied only when comparing our results to 2MASS or in comparing to 2MTF. The 2MASS magnitudes ($M^{\prime}$) are first transformed to the SIRIUS system ($M$) via the following colour-dependent transformation equations \citetext{Nakajima priv. comm.}:
\[
 J = J^{\prime} + (-0.045\pm0.008)(J-H)^{\prime} - (0.001\pm0.008) 
\]
\[
H = H^{\prime} + (0.027\pm0.007)(J-H)^{\prime} -(0.009\pm0.008)
\]
\[
 K_{s} = K_{s}^{\prime} +(0.015\pm0.008)(J-K_s)^{\prime} -(0.001\pm0.008)
\]
which are valid over a wide range in colour $0.0 \le (J-H) \le 3.9$, $0.0 \le (H-K_s) \le 2.8$ and $0.0 \le (J-K_s) \le 4.9$.  

\begin{figure}
\centering
\includegraphics[width=0.5\textwidth]{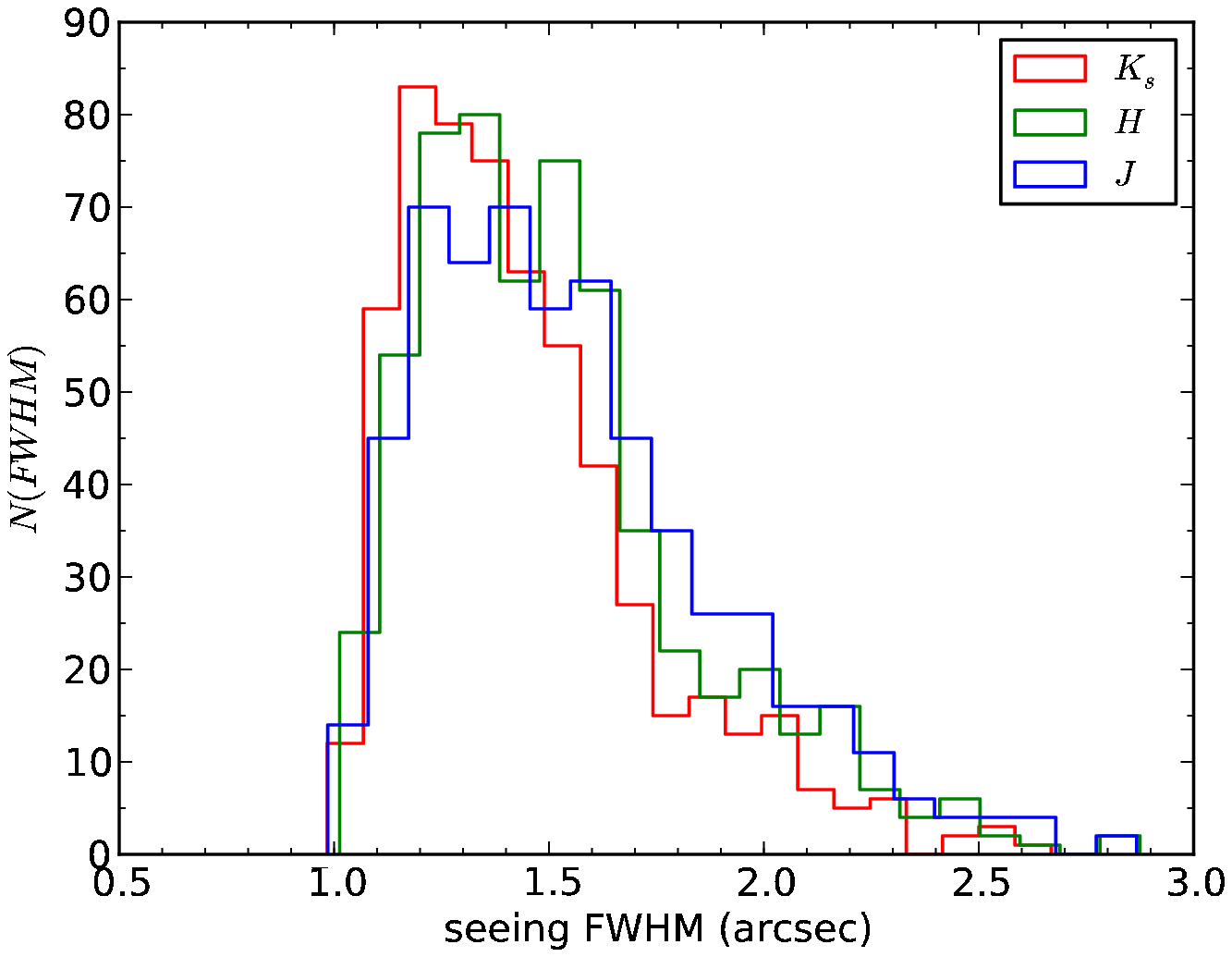}
\includegraphics[width=0.5\textwidth]{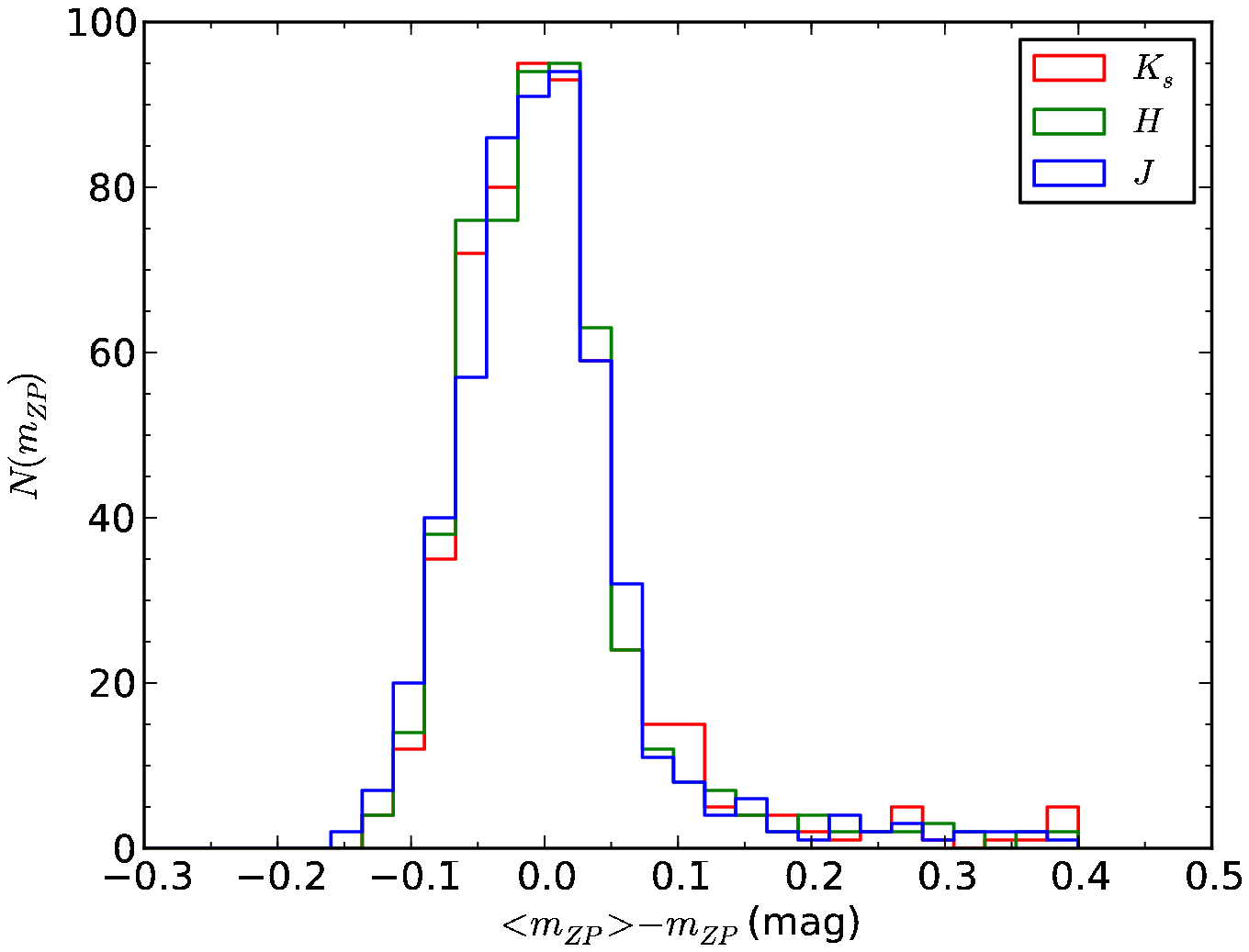}
 \caption{Histograms of  image FWHM as a measure of the seeing and image magnitude zero points as a measure of the image quality.}
 \label{fig:image_hist}
\end{figure}

Two measures of the photometric conditions are determined for each field: the seeing, or FWHM of stars in the field, and the magnitude zero-point. Figure~\ref{fig:image_hist} shows the distribution of seeing values and magnitude zero-points for all of the final $578$ images. $65$~per~cent of the images were observed with $K_s$ band seeing of  $< 1\farcs5$ and 85~per~cent $ < 1\farcs75$. The median zero-points are consistent with values previously determined by \cite{Riad2010} and \cite{Cluver+2008}. The medians and standard deviations of both the magnitude zero-points and seeing values in all three bands are given in Table~\ref{tab:imstat}.

\begin{table}
 \centering
\caption{Mean magnitude zero-points and average seeing for all $578$ images}
\label{tab:imstat}
 \begin{center}
\begin{tabular}{@{}ccccc@{}}
\hline
 & $<ZP>$ & $\sigma$ & seeing & $\sigma$ \\ 
 & [mag] & [mag] & [arcsec] & [arcsec] \\ 
\hline
$J$ & 20.84 & 0.06 & 1.49 & 0.35 \\ 
$H$ & 21.02 & 0.06 & 1.47 & 0.33 \\ 
$K_s$ & 20.24 & 0.06 & 1.39 & 0.30 \\
\hline
 \end{tabular}
 \end{center}
\end{table}

\section{Surface Photometry}
\label{sect:phot}
\subsection{Galaxy Identification}
Possible \reply{NIR counterparts to the \HI~sources} were identified in the respective NIR images by a
visual search of the \reply{false} three-colour \reply{(RGB)} images generated from the \reply{$K_s$, $H$ and $J$ bands where the $K_s$ image is mapped to the red channel, the $H$ image to green and and the $J$ image to blue}.  Their different colour and extended nature allows
galaxies to be readily identified by eye. A \textsc{python} script, that uses the \textsc{pyDS9}\footnote{http://hea-www.harvard.edu/saord/ds9/pyds9/} interface with SAOImage \textsc{DS9}\footnote{This research has made use of SAOImage DS9, developed by Smithsonian Astrophysical Observatory.}, was used to interactively and systematically search the $7.7\farcm \times 7\farcm7$ images for possible \reply{counterparts to the \HI~sources}.  The likelihood of a counterpart was assessed based on the NIR images and the \HI~spectral features. For example, an obvious elliptical galaxy would be excluded in preference of an edge-on spiral galaxy if the \HI~profile showed a clear double-horn structure. This is described further in Sect.~\ref{sect:nircntparts}.

\subsection{Star-subtraction}
The increase in stellar density near the Galactic plane results in heavy contamination by foreground stars. Star-subtraction by fitting the point spread function (PSF) was employed to remove the flux contribution of the foreground stars from the galaxy flux.  This is implemented in two steps:  the PSF is first determined for the field, which may be variable across the field. Second, the fitted PSF is used to remove stars in the neighbourhood of each detected galaxy. The automated PSF fitting routine for the Norma Wall Survey \citetext{Nagayama priv. comm.} was  modified to improve the subtraction of stars on edge-on disks and to prevent the removal of sub-structure within the disks of face-on spirals. The method of star subtraction is based on the KILLALL routine \citep{Buta&McCall1999}. The steps, implemented in \textsc{pyRAF}, are: 
\begin{enumerate}
 \item The sky background and rms, $\sigma$, in the image is determined using \textsc{imstatistics} with 30 iterations with $3\sigma$ clipping.
 \item \label{step:model} The galaxy is modelled using \textsc{ellipse} and \textsc{bmodel}. This model is subtracted from the image resulting in the galaxy-subtracted image. Structures on the galaxy, in particular spiral arms, may not be fully modelled and may result in residuals in the  galaxy-subtracted image.
 \item The bright stars (above $3.5\sigma$ of the background) in the galaxy-subtracted image are detected and removed with \textsc{SExtractor} \citep{BertinArnouts1996}. The \textsc{daophot} tasks \textsc{phot} and \textsc{allstar} are used to measure the PSF photometry of these sources and then to remove them from the galaxy-subtracted image.
 \item The faint stars (above $1.8\sigma$) are then detected and removed from the galaxy-subtracted image. Within the radius of the galaxy only stars that are also detected by \textsc{daofind} with a threshold of $2\sigma$ are included in the list. This prevents the misidentification of residuals of structures as stars. Again PSF photometry is done with \textsc{phot} and the stars are removed from the galaxy-subtracted image using \textsc{allstar}.
 \item The two \textsc{allstar} output files for the bright and faint stars are concatenated, and used in \textsc{substar} to remove these stars from the original image.
 \item \label{step:res}Any residuals as a result of the imperfect fit of the PSF to every star in the image are found and removed.
\end{enumerate}
Steps \ref{step:model} to \ref{step:res} are iterated four times. Each iteration improves the galaxy model because the subtraction of stars on the galaxy gets more accurate. This results in improved photometry of stars on the galaxy and therefore increasingly reliable subtraction.
\begin{figure*}
\begin{center}
\includegraphics[width=\textwidth]{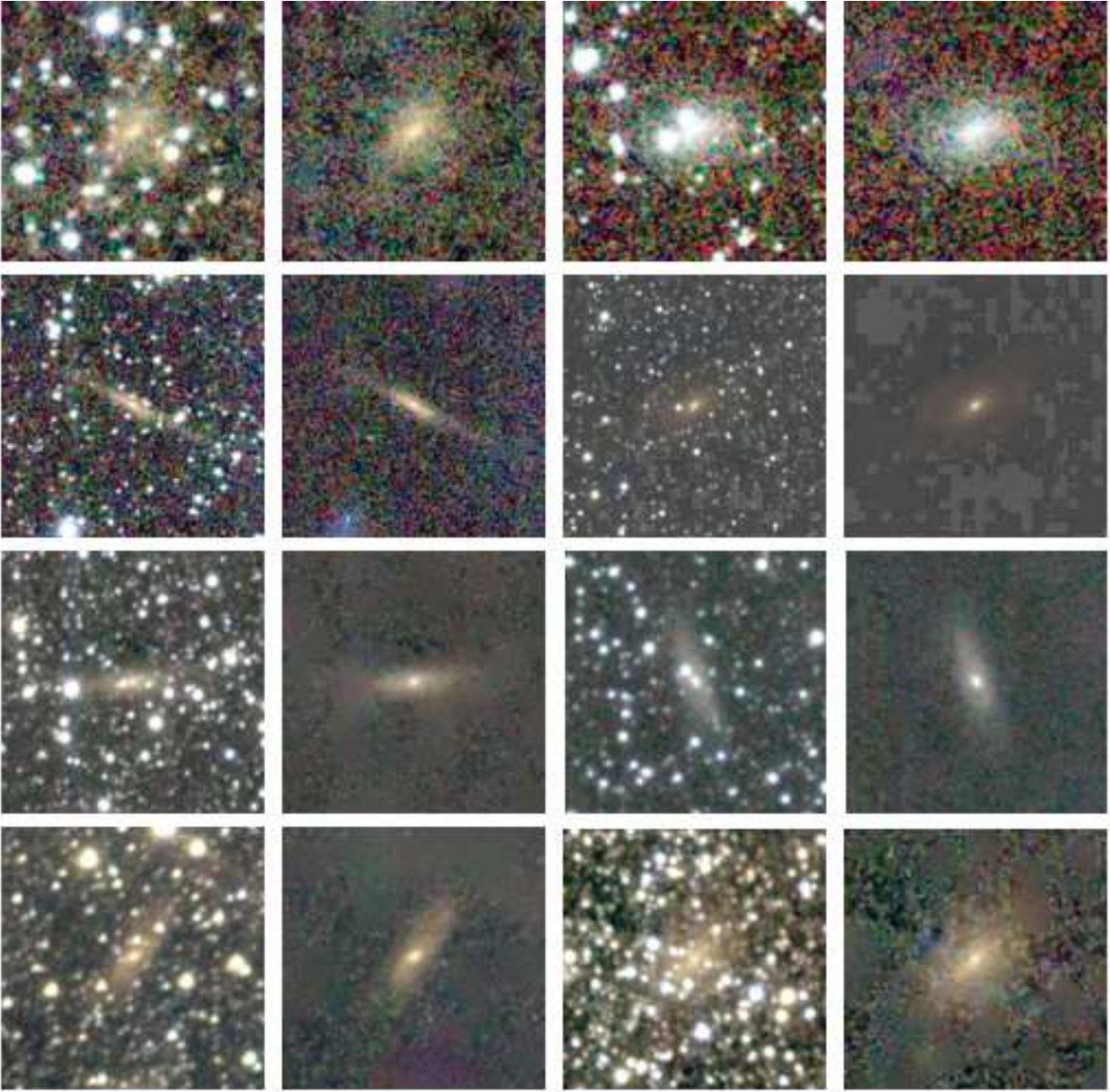}
 \caption{Examples of starry and star-subtracted postage stamp images. The colours of the galaxies are representative of the observed NIR colours.}
 \label{fig:starsub}
\end{center}
\end{figure*}

Figure~\ref{fig:starsub} shows some examples of the results of the star-subtraction routine. It displays qualitatively that the star-subtraction performs well. The removal of stars in this manner has been shown to have a minimal effect on the uncertainty in the galaxy photometry. In simulations of subtracting an artificially added star-field from galaxies, \cite{Nagayama2004} found that the measured ${K_s}_{20}$ magnitudes (the magnitude within the radius to the $K_s = 20$ \marc isophote)  were on average $0.04$~mag fainter with a standard deviation of $0.20$~mag.

\subsection{Astrometric Parameters}
\subsubsection{Source Positions}
\label{sect:pos}
The position is measured using the intensity-weighted centroid of the $J+H+K_s$ ``super'' co-add image. This position should be more precise as it measures the centroid from higher signal-to-noise data. The catalogue identifiers are \texttt{RA}, \texttt{Dec} for the  J2000 Equatorial coordinates  in degrees and \texttt{l}, \texttt{b} for the  Galactic coordinates in degrees.

\subsubsection{Ellipse Fitting and Object Orientation}
Ellipses were fitted to each galaxy image using the \textsc{iraf} task \textsc{ellipse}. The central coordinates, ($X,Y$), are fixed while  the position angle ($\phi$, measured counter-clockwise from the $x$-axis) and ellipticity ($\epsilon = 1-b/a$) are fitted at intervals in the semi-major axis. \textsc{ellipse} returns a table containing, amongst others, the intensity in counts enclosed within the ellipse, the position angle, the ellipticity and their errors at each semi-major axis step. The ellipticity and position angle characteristic of the galaxy are determined to be the average value in the outer part of the disk, between the $1\sigma$ and $2\sigma$ isophotes, where $\sigma$ is the sky rms. Both the ellipticity and the position angle are usually stable in the outer disk. The ellipse parameters are determined individually in all three bands. 

The ellipticities are identified in the catalogue as \texttt{<band>\_ell} where \texttt{<band>} is one of \texttt{j}, \texttt{h}  or \texttt{k} for the $J$, $H$ or $K_s$ bands respectively. The $1\sigma$ errors are identified by \texttt{<band>\_ell\_sig}. The position angles and errors (in  J2000 coordinates, measured East of North) are listed in the catalogue as \texttt{<band>\_phi} and \texttt{<band>\_phi\_sig} respectively.

\subsection{SELF-sky Correction for Large Galaxies}
The nature of the \textit{self-sky} measurement means that galaxies larger than the dither radius overlap in the individual dithered frames. The final sky frame therefore contains an extended residual, typically of a few counts per pixel above the average sky value, in the region of the galaxy. The result is that too much sky is subtracted from the region around the galaxy causing the measured fluxes to be lower. This is clearly evident in the intensity profile of the galaxy as a dip in flux around the outer edges below the median sky level. The residual is such that it is largest in the centre of the galaxy, decreasing towards the outer parts. However, it cannot be measured directly at the centre of the galaxy. We therefore measure the shape of the recovery of the dip in the outer parts to the sky level and extrapolate to the centre. This is done by fitting a linear function to the intensity profile between the minimum intensity and the asymptotic sky value and adding this to the intensity profile. A linear function was chosen as it is the simplest function that describes the shape of the dip. It provides a first order correction while introducing as little error as possible from higher order terms. The size of the correction at the centre of the galaxy can be up to $5-10$\ counts\,pixel$^{-1}$ for large galaxies with bright extended cores. However, this affects only a small number of galaxies, $\sim 20$.

\begin{figure}
\centering
 \includegraphics[width=0.5\textwidth]{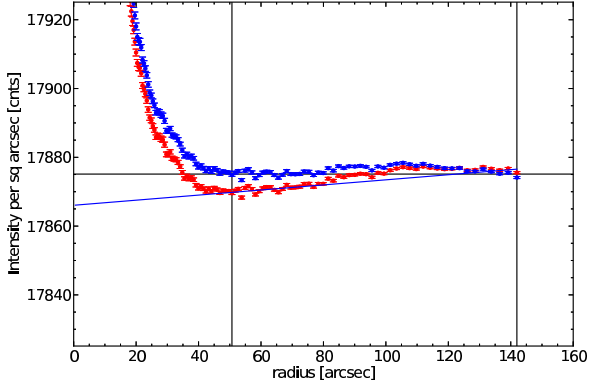}
\caption[Intensity profile showing SELF-sky correction]{Example intensity profile with SELF-sky correction. The red profile shows the intensity profile before the correction. A linear function is fitted to the outer part of the profile from the minimum (\reply{blue} line) to the outer edge where the sky level is reached. The limits of the fit are shown with solid vertical lines. The horizontal line shows the measured median sky level in the image. The blue profile shows the corrected profile.}
 \label{fig:selfcor}
\end{figure}

A correction for this effect is made for galaxies which have both semi-major and -minor axes that are larger than the dither radius. As an example for one galaxy (J0748-26B), Fig.~\ref{fig:selfcor} shows the $H$ band intensity profile. The red profile is uncorrected. The \reply{blue} line shows the fit that has been made from the minimum in the profile to the outer edges. The fit region is demarcated by two vertical lines. The horizontal line shows the measured median sky value. The blue profile shows the corrected intensity profile.

\subsection{Photometric Parameters}

\subsubsection{Fixed Aperture Magnitudes}
The simplest photometric measurement is for fixed circular apertures. Magnitudes are measured using a \textsc{python} routine for apertures of $3''$, $5''$, $7''$, $10''$ and $15''$. The magnitudes and their errors are identified in the catalogue as \texttt{<band>\_m<c>}, \texttt{<band>\_msig<c>}, where \texttt{<band>} is the band in which the magnitude is measured and \texttt{<c>} is the radius, in arcsec, of the aperture.

\subsubsection{Radial Surface Brightness Profiles}
The one-dimensional radial surface brightness profile (SBP) is measured with \texttt{ellipse} by holding the centre,  position angle and ellipticity fixed. The SBPs of each galaxy were fitted with a double S\'{e}rsic function,
\[
 \mu(a) = \mu_0 +1.086\left(\frac{a}{r_h}\right)^n + {\mu_0}_{2} +1.086\left(\frac{a}{r_{h_2}}\right)^{n_2},
\]
or in terms of intensities,	
\[
 I(a) = I_0 \exp\left(\frac{a}{r_h}\right)^n + {I_0}_{2} \exp\left(\frac{a}{r_{h_2}}\right)^{n_2}.
\]
The double S\'{e}rsic function is the most general profile that allows for the simultaneous fitting of both an inner bulge (usually following a de Vaucouleurs profile with $n=4$) and an outer disk profile (exponential with $n_2=1$).

\begin{figure*}
\centering
 \includegraphics[width = 0.33\textwidth]{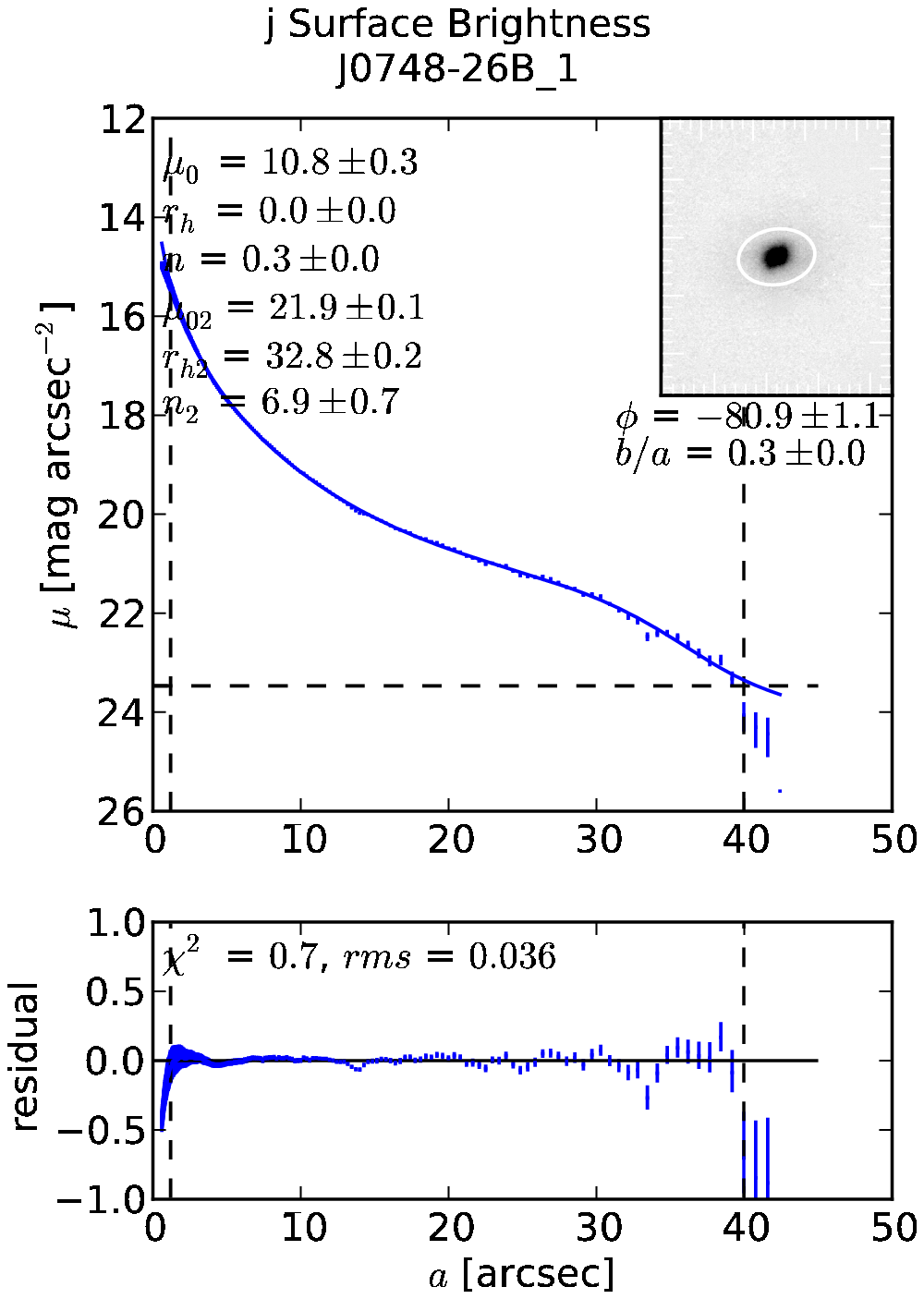}
 \includegraphics[width = 0.33\textwidth]{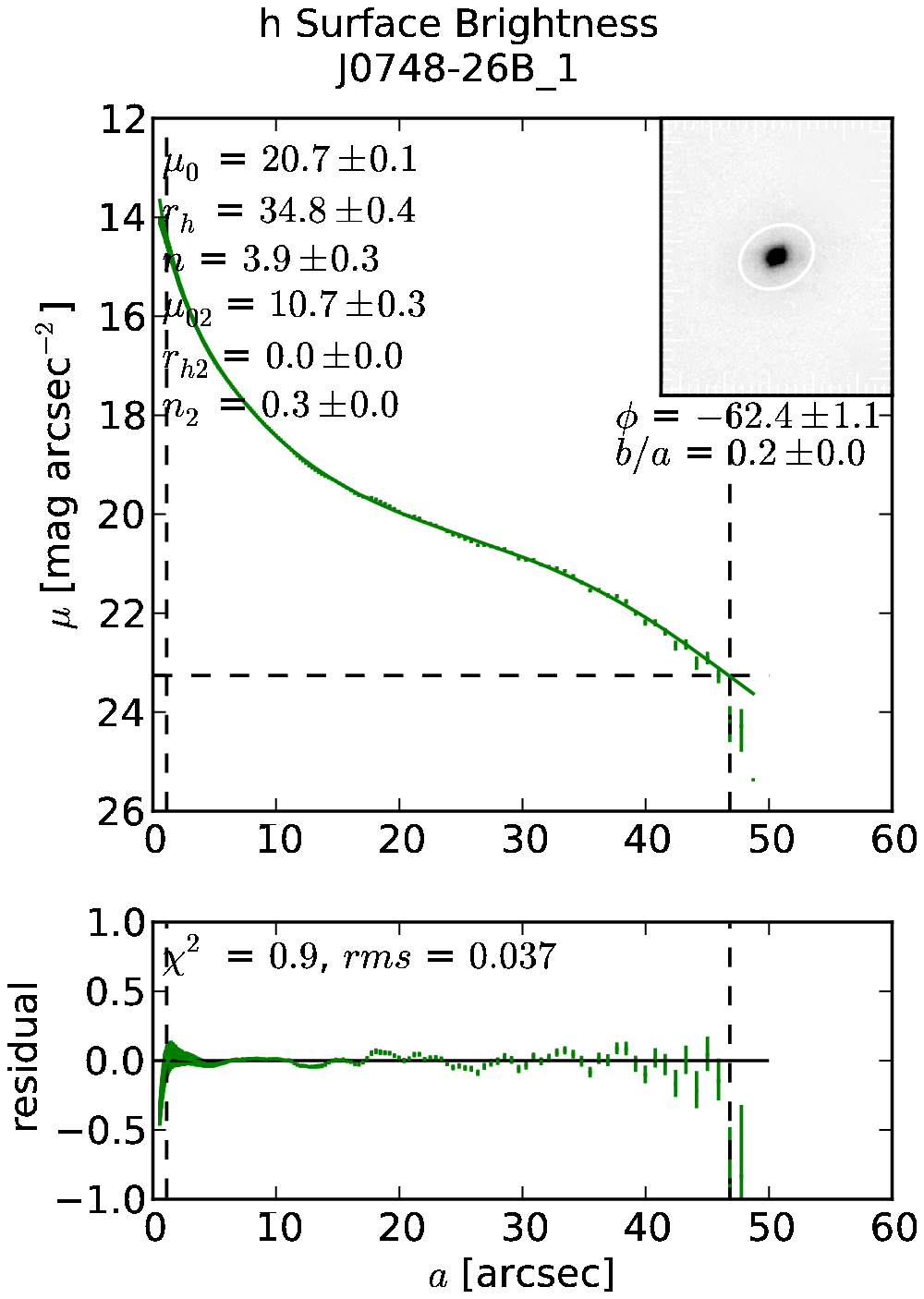}
 \includegraphics[width = 0.33\textwidth]{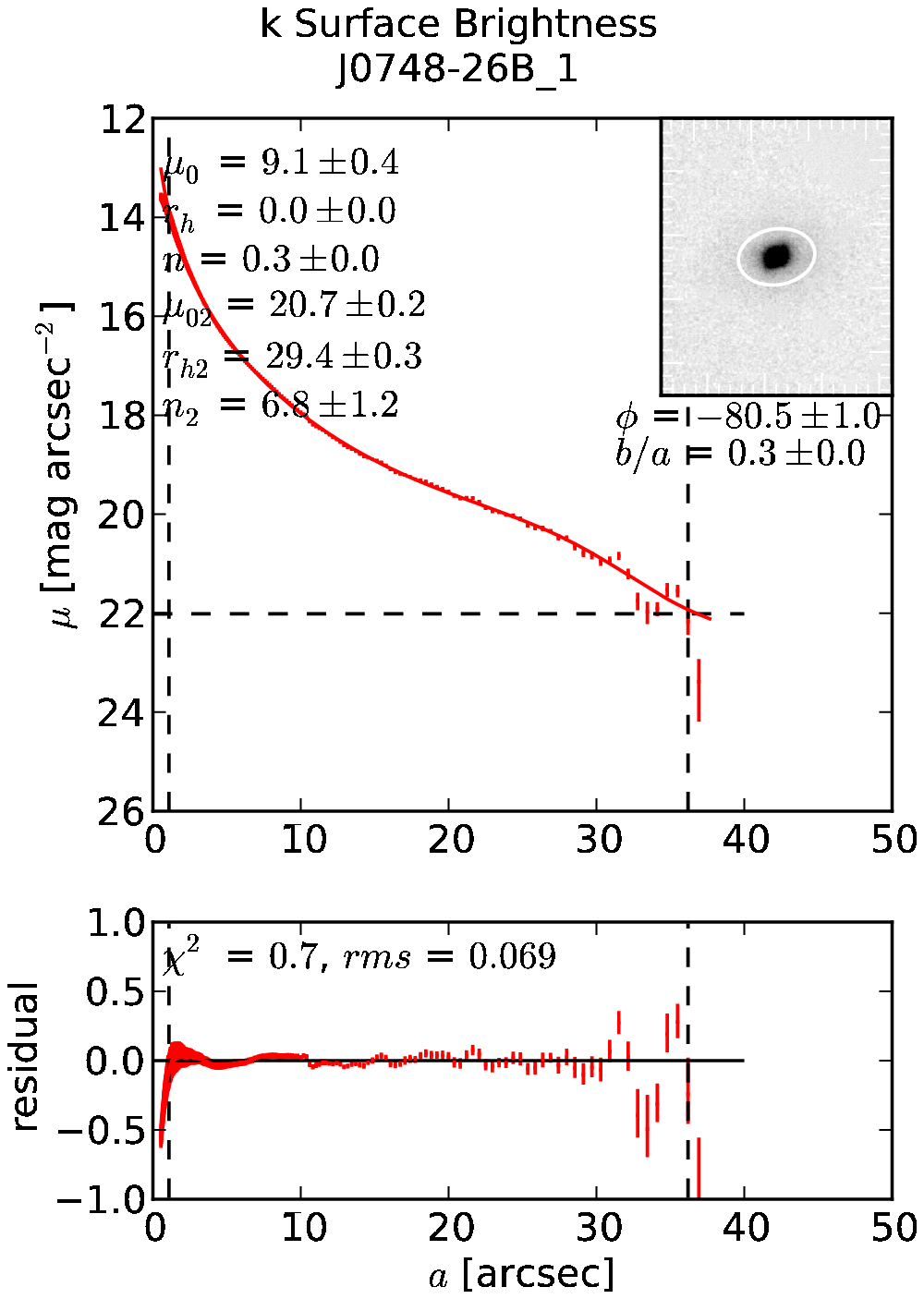}
 \caption[Example of Radial SBP plots and fits]{Example of radial SBP plots in the $J$ (\textit{left}), $H$ (\textit{centre}) and $K_s$ (\textit{right}) bands for galaxy J0748-26B. The best-fit double S\'{e}rsic function to the outer part of the profile is shown as a solid line together with the residuals in the bottom panel. The rms of the fit, in mag\,arcsec$^{-2}$, is given in the top left corner of the bottom panel. The inset panel shows a greyscale image of the galaxy, with the characteristic ellipse in white. The parameters of the ellipse are given below the inset and the radius of the ellipse is determined to be the $1\sigma$ isophote, which is marked on the SBP by the horizontal and vertical dashed lines.}
\label{fig:SBP}
\end{figure*}

A plot of the SBP in each band is produced which shows the measured SBP, the fitted SBP, the residuals, the fitted parameters, a grey-scale image and the ellipse parameters. Examples of such SBP plots for J0748-26B are shown in Fig.~\ref{fig:SBP}.

\subsubsection{Isophotal Magnitudes}
Isophotal elliptical aperture photometry is performed in each band, using the \texttt{ellipse} task. The isophotal radius, $r_{\mu}$, is the radius of the isophote at surface brightness $\mu$. The isophotal magnitude, $m_{\mu}$, is defined as the integrated magnitude within that radius. In each band the $\mu = 20$~mag~arcsec$^{-2}$, $\mu = 21$~\marc and $\mu = 22$~\marc isophotal radii and magnitudes are calculated. The surface brightness profile is used to determine the radius at which a given surface brightness is reached. The catalogue names are \texttt{r\_<band>20e}, \texttt{<band>\_m\_<band>20e} and \texttt{<band>\_msig\_<band>20e} for the $20$~mag~arcsec$^{-2}$ isophotal radius, magnitude and error respectively in the given  band. The fiducial isophotal parameters are \texttt{r\_<band>20fe}, \texttt{<band>\_m\_<fband>20fe} and \texttt{<band>\_msig\_<fband>20fe} for the $20$~mag~arcsec$^{-2}$ isophotal radius, magnitude and error respectively in the given  band where the fiducial band is \texttt{fband}.

\subsubsection{Total Magnitudes}
Following a similar analysis to \cite{Kirby+2008}, extrapolation of the analytic double S\'{e}rsic function allows for an accurate estimate of the flux that remains undetected below the sky noise. The integrated flux between a given outer radius, $r_{\rm{out}}$, and $r=\infty$  is:
\[
 \Delta m = -2.5 \log \left( \frac{I_{\rm{missing}}}{I_{\rm{total}}} \right) 
\]
where
\[
 \nonumber I_{\rm{missing}} =  \int_{0}^{a_{\rm{tot}}} 2\pi (1-\epsilon) I(a) a \,da
\]
\[
 = \int_{0}^{a_{\rm{tot}}} 2\pi (1-\epsilon) \left[I_0 \exp\left(\frac{a}{r_h}\right)^n + {I_0}_2 \exp\left(\frac{a}{r_{h_2}}\right)^{n_2} \right] a\, da
\]
\[
 = 2 \pi (1-\epsilon) \left\lbrace \frac{ I_0 r_h^2}{n} \Gamma\left[ \frac{2}{n} , \left(\frac{r_{\rm{tot}}}{r_h} \right)^n \right] + \frac{ {I_0}_2 r_{h_2}^2}{n_2} \Gamma\left[ \frac{2}{n_2} , \left(\frac{r_{\rm{tot}}}{r_{h_2}} \right)^{n_2} \right] \right\rbrace 
\]
and
\[
I_{\rm{total}}=  \int_{0}^{\infty} 2\pi (1-\epsilon) I(a) a \,da
\]
\[
 = \int_{0}^{\infty} 2\pi (1-\epsilon) \left[ I_0 \exp\left(\frac{a}{r_h}\right)^n + {I_0}_{2} \exp\left(\frac{a}{r_{h_2}}\right)^{n_2}\right] a\,  da 
\]
\[
= 2 \pi (1-\epsilon) \left\lbrace \frac{ I_0 r_h^2}{n} \Gamma\left[ \frac{2}{n} \right] + \frac{ {I_0}_{2} r_{h_2}^2}{n_2} \Gamma\left[ \frac{2}{n_2} \right] \right\rbrace .
\]
$\Gamma [\alpha,x ]$ is the incomplete gamma function and $a_{\rm{tot}}$ was taken to be the radius at which the surface brightness drops to $1\sigma$. The observed total magnitude, $m_{\rm{obs}}$, is the magnitude enclosed within this aperture. The total magnitude is then determined to be:
\[
 m_{\rm{tot}} = m_{\rm{obs}} - \Delta m.
\]
                                                                                                                                                    
The catalogue parameters are \texttt{<band>\_m\_ext} for the extrapolated total magnitude, \texttt{<band>\_msig\_ext} for the error error [mag] and \texttt{<band>\_dm\_ext} for the difference between total magnitude and the measured outer magnitude.

\subsubsection{Photometric Errors}
The fundamental limits to the photometry depend on  the accuracy of the background  determination, the $S/N$ within the aperture and the zero-point calibration. The error, $\sigma_F$, on a flux within a given aperture, $F$, is the quadrature sum of the Poissonian source noise and the error in the sky background:
\[
 \sigma_F = \sqrt{F +N \sigma_{rms}^2},
\]
where $N$ is the number of pixels in the aperture and $\sigma_{rms}$ is the sky background

The ellipse fitting procedures will introduce additional errors. However, these are neglected here as they are expected to be less significant than the errors already mentioned.

 \reply{
\subsection{Galaxy Types}
\label{sect:types}
Based on visual inspection of the NIR images we made an estimate of the morphological type of each galaxy: galaxies with no obvious bulge were classed as late-type (Hubble T-type 5), those with a clear bulge were classed as early-type spirals (T-type 1) and those with some evidence of a bulge as intermediate (T-type 3). Consideration was made for the effects of extinction, which can lead to the outer parts of the galaxies disappearing and the galaxy appearing more `bulge'-like, however, we do note that this is an estimate only. Approximately $15$~per~cent were classed as early type spirals, $70$~per~cent were classed as intermediate type spirals and $15$~per~cent were classed as late type spirals.
}

\subsection{Catalogue}
The $40$ brightest galaxies in the catalogue are presented, as an example, in Table~\ref{tab:cat}. The columns are as follows: Column 1 -- Unique identifier [ZOAhhmmss.sss$\pm$ddmmss.ss] from the co-add centroid (see Sect.~\ref{sect:pos}); Column 2 -- \HI~catalogue name [Jhhmm$\pm$dd]; Column 3 -- Counterpart Number (can be greater than 1 where more than one possible counterpart was identified in the IRSF field of view) followed by a $*$ if it is confirmed; Columns 4 \& 5 --  Galactic co-ordinates [deg]; Columns 6 -- $J$ band ellipticity ($\epsilon = 1 - b/a$) and error; Column 7 --  $J$ band  position angle (East of North) and error [deg]; Column 8 --   ${K_s}_{20}$ fiducial isophotal radius [arcsec]; Columns 9--11 -- $J$, $H$ and $K_s$ band  ${K_s}_{20}$ fiducial isophotal magnitude and error [mag]; Columns 12--14 --   $J$, $H$ and $K_s$ band extrapolated total magnitude [mag]; Column 15 --  Galactic reddening along the line of sight \citep{Schlegel1998} [mag]; and Column 16 --  Stellar density for stars brighter than $14$~mag in $K_s$ detected with the IRSF. All magnitudes are given as IRSF/SIRIUS magnitudes, not transformed to 2MASS. The catalogue in its entirety, with all measured parameters,  is available in the electronic supplement.

\subsection{Comparison with 2MASX}
\label{sect:comp2mass}
To evaluate the consistency of our photometry we compare the astrometric and photometric parameters of a sample of our galaxies that appear in the 2MASX catalogue \citep{Jarrett+2000a}. 2MASX positional errors are typically around $0\farcs5$. Sources are assumed to be identical in the two catalogues if they are separated by less than $1\farcs5$. This small correlation radius is taken because the SIRIUS images have been astrometrically calibrated against the 2MASX. Moreover, in many cases, the 2MASS coordinates are offset from the IRSF coordinates because the source is centred on a nearby foreground star which has not been removed. Only sources with reliable IRSF photometry are compared, i.e. galaxies near the edge of the field are excluded; $102$ galaxies are included in the comparison.


\begin{landscape}
 \begin{table}
 \centering
  \caption{Selected measured galaxy parameters for the forty brightest ${K_s}_{20}$ magnitude galaxies in the catalogue.}
   \label{tab:cat}
   \begin{small}
  \begin{tabular}{lllrrrrrr@{\,$\pm$\,}lr@{\,$\pm$\,}lr@{\,$\pm$\,}lr@{\,$\pm$\,}lr@{\,$\pm$\,}lr@{\,$\pm$\,}l@{}c@{}r}
  \hline
Designation	&	\HI~Name	& \# &		\multicolumn{1}{c}{$l$}		&		\multicolumn{1}{c}{$b$}	&	\multicolumn{1}{c}{$\epsilon_J$}	&		\multicolumn{1}{c}{$\phi_{J}$}	&	\multicolumn{1}{c}{$r_{k20fe}$}	&	\multicolumn{2}{c}{$J_{k20fe}$}	&	\multicolumn{2}{c}{$H_{k20fe}$}	&	\multicolumn{2}{c}{${K_s}_{k20fe}$}	&	\multicolumn{2}{c}{$J_{tot}$}	&	\multicolumn{2}{c}{$H_{tot}$}	&	\multicolumn{2}{c}{${K_s}_{tot}$}	&	\multicolumn{1}{c}{$E(B$-$V)$}	&	\multicolumn{1}{c}{$SD$}\\
~	&	~	& &		\multicolumn{1}{c}{[deg]}		&		\multicolumn{1}{c}{[deg]}	&	\multicolumn{1}{c}{~}	&		\multicolumn{1}{c}{[deg]}	&	\multicolumn{1}{c}{[$''$]}	&	\multicolumn{2}{c}{[mag]}	&	\multicolumn{2}{c}{[mag]}	&	\multicolumn{2}{c}{[mag]}	&	\multicolumn{2}{c}{[mag]}	&	\multicolumn{2}{c}{[mag]}	&	\multicolumn{2}{c}{[mag]}	&	\multicolumn{1}{c}{[mag]}	&	\multicolumn{1}{c}{~}\\
\hline
ZOA141309.873-652020.76 & J1413-65 & 1$^*$ & 311.326 & -3.808 & 0.44 & 37.6 & 87.05 & 7.50 & 0.02 & 6.49 & 0.02 & 6.09 & 0.02 & 7.08 & 0.12 & 6.39 & 0.02 & 6.17 & 0.03 & 1.47 & 4.37\\
ZOA151434.147-525921.52 & J1514-53 & 1$^*$ & 323.594 & 4.043 & 0.76 & -16.9 & 136.09 & 8.86 & 0.02 & 7.91 & 0.02 & 7.42 & 0.02 & 8.95 & 0.03 & 7.83 & 0.03 & 7.34 & 0.02 & 0.99 & 4.30\\
ZOA085728.494-391605.86 & J0857-39 & 1$^*$ & 261.500 & 4.100 & 0.09 & 27.6 & 44.85 & 9.19 & 0.02 & 8.32 & 0.02 & 7.97 & 0.02 & 9.20 & 0.02 & 8.09 & 0.06 & 8.02 & 0.03 & 0.72 & 3.61\\
ZOA150928.962-523320.67 & J1509-52 & 1$^*$ & 323.155 & 4.810 & 0.79 & 45.3 & 98.73 & 9.24 & 0.02 & 8.38 & 0.02 & 7.97 & 0.02 & 8.77 & 0.14 & 8.35 & 0.03 & 7.86 & 0.03 & 0.77 & 4.09\\
ZOA122238.290-583657.66 & J1222-58 & 1$^*$ & 299.180 & 4.046 & 0.69 & -27.0 & 60.30 & 9.71 & 0.02 & 8.92 & 0.02 & 8.60 & 0.02 & 9.68 & 0.02 & 8.57 & 0.10 & 8.61 & 0.02 & 0.58 & 4.12\\
ZOA145709.815-542331.46 & J1457-54 & 1$^*$ & 320.654 & 4.096 & 0.50 & 51.6 & 53.81 & 10.05 & 0.02 & 9.14 & 0.02 & 8.69 & 0.02 & 9.83 & 0.08 & 8.94 & 0.06 & 8.59 & 0.02 & 0.85 & 4.21\\
ZOA094916.505-475511.27 & J0949-47A & 1 & 274.257 & 4.549 & 0.06 & 28.9 & 38.65 & 9.67 & 0.02 & 8.98 & 0.02 & 8.69 & 0.02 & 9.59 & 0.04 & 8.85 & 0.04 & 8.71 & 0.03 & 0.35 & 3.80\\
ZOA114606.371-562326.95 & J1145-56 & 1$^*$ & 293.937 & 5.336 & 0.27 & -48.5 & 54.68 & 9.99 & 0.02 & 9.26 & 0.02 & 8.82 & 0.02 & 9.41 & 0.20 & 8.45 & 0.24 & 8.41 & 0.11 & 0.39 & 3.84\\
ZOA135138.534-583515.22 & J1351-58 & 1$^*$ & 310.724 & 3.370 & 0.52 & 32.3 & 47.27 & 10.23 & 0.02 & 9.31 & 0.02 & 8.87 & 0.02 & 10.18 & 0.03 & 9.27 & 0.02 & 8.55 & 0.09 & 0.97 & 4.29\\
ZOA080610.996-273140.54 & J0805-27 & 1 & 245.709 & 2.414 & 0.05 & -13.3 & 52.34 & 10.39 & 0.02 & 9.38 & 0.02 & 8.89 & 0.02 & 10.45 & 0.03 & 9.37 & 0.02 & 8.86 & 0.02 & 0.42 & 3.78\\
ZOA074752.048-184453.18 & J0747-18 & 1$^*$ & 236.009 & 3.374 & 0.82 & -82.1 & 84.90 & 10.06 & 0.02 & 9.31 & 0.02 & 8.91 & 0.02 & 9.83 & 0.06 & 8.80 & 0.14 & 8.64 & 0.08 & 0.38 & 3.55\\
ZOA143158.829-552758.82 & J1431-55 & 1$^*$ & 316.912 & 4.653 & 0.05 & 82.6 & 34.39 & 10.32 & 0.02 & 9.46 & 0.02 & 9.05 & 0.02 & 9.74 & 0.19 & 8.60 & 0.27 & 8.56 & 0.15 & 0.85 & 4.09\\
ZOA163211.878-280530.82 & J1632-28 & 1$^*$ & 351.084 & 13.502 & 0.72 & -52.8 & 81.48 & 10.28 & 0.02 & 9.45 & 0.02 & 9.10 & 0.02 & 10.13 & 0.06 & 9.28 & 0.03 & 8.97 & 0.02 & 0.61 & 3.54\\
ZOA141036.181-653457.76 & J1410-65 & 1$^*$ & 310.997 & -3.958 & 0.52 & -87.4 & 46.18 & 10.35 & 0.02 & 9.48 & 0.02 & 9.14 & 0.02 & 10.31 & 0.03 & 9.39 & 0.02 & 8.72 & 0.12 & 0.60 & 4.32\\
ZOA074843.902-261446.34 & J0748-26 & 1$^*$ & 242.586 & -0.239 & 0.30 & -27.9 & 31.93 & 10.71 & 0.02 & 9.78 & 0.02 & 9.25 & 0.02 & 9.82 & 0.29 & 9.70 & 0.03 & 8.97 & 0.08 & 0.72 & 3.84\\
ZOA091645.231-542124.17 & J0916-54B & 2 & 274.871 & -3.630 & 0.67 & 56.3 & 30.08 & 9.61 & 0.02 & 9.27 & 0.02 & 9.28 & 0.02 & 8.84 & 0.25 & 8.70 & 0.17 & 8.83 & 0.13 & 0.94 & 3.99\\
ZOA083439.531-400855.61 & J0834-40 & 1$^*$ & 259.448 & 0.122 & 0.22 & 20.8 & 34.38 & 10.02 & 0.02 & 9.52 & 0.02 & 9.30 & 0.02 & 9.41 & 0.21 & 9.11 & 0.12 & 8.55 & 0.22 & 2.12 & 3.84\\
ZOA132723.827-572922.23 & J1327-57 & 1$^*$ & 307.768 & 5.044 & 0.74 & 1.6 & 46.73 & 10.86 & 0.02 & 9.82 & 0.02 & 9.30 & 0.02 & 11.03 & 0.02 & 9.76 & 0.02 & 9.10 & 0.05 & 0.81 & 3.98\\
ZOA085838.816-423157.53 & J0858-42 & 1$^*$ & 264.125 & 2.141 & 0.44 & 63.0 & 42.93 & 12.14 & 0.02 & 10.09 & 0.02 & 9.35 & 0.02 & 12.23 & 0.04 & 9.75 & 0.08 & 9.20 & 0.04 & 3.99 & 3.73\\
ZOA094952.868-563235.55 & J0949-56 & 1$^*$ & 279.808 & -2.054 & 0.24 & -67.6 & 45.57 & 11.43 & 0.02 & 10.29 & 0.02 & 9.58 & 0.02 & 11.03 & 0.23 & 9.78 & 0.12 & 9.01 & 0.14 & 2.10 & 4.16\\
ZOA141933.720-580850.19 & J1419-58 & 1$^*$ & 314.363 & 2.755 & 0.59 & 19.7 & 52.89 & 11.00 & 0.02 & 10.14 & 0.02 & 9.59 & 0.02 & 10.80 & 0.08 & 10.02 & 0.03 & 9.41 & 0.03 & 1.52 & 4.36\\
ZOA163140.118-280606.66 & J1631-28 & 1$^*$ & 350.997 & 13.584 & 0.27 & 69.5 & 35.29 & 10.83 & 0.02 & 10.07 & 0.02 & 9.59 & 0.02 & 10.70 & 0.05 & 9.40 & 0.20 & 9.45 & 0.03 & 0.63 & 3.60\\
ZOA090033.130-392627.15 & J0900-39 & 1$^*$ & 262.020 & 4.438 & 0.40 & -53.9 & 37.84 & 10.79 & 0.02 & 10.15 & 0.02 & 9.67 & 0.02 & 10.70 & 0.03 & 10.18 & 0.03 & 9.63 & 0.02 & 0.67 & 3.61\\
ZOA161319.695-562349.17 & J1613-56 & 1$^*$ & 328.261 & -3.802 & 0.35 & -70.7 & 40.49 & 10.93 & 0.02 & 10.69 & 0.02 & 9.68 & 0.02 & 10.58 & 0.13 & 9.82 & 0.31 & 9.29 & 0.09 & 0.54 & 4.43\\
ZOA162101.624-360831.49 & J1621-36 & 1$^*$ & 343.413 & 9.765 & 0.55 & 54.2 & 38.44 & 10.84 & 0.02 & 10.05 & 0.02 & 9.71 & 0.02 & 10.51 & 0.11 & 9.83 & 0.05 & 9.41 & 0.08 & 0.71 & 3.79\\
ZOA074252.007-315959.79 & J0742-31 & 1$^*$ & 246.932 & -4.226 & 0.52 & 31.9 & 26.61 & 11.15 & 0.02 & 10.38 & 0.02 & 9.71 & 0.02 & 11.08 & 0.04 & 10.41 & 0.02 & 9.33 & 0.11 & 0.83 & 3.76\\
ZOA101212.032-623159.40 & J1012-62 & 1$^*$ & 285.685 & -5.121 & 0.62 & 41.4 & 39.84 & 10.84 & 0.02 & 10.08 & 0.02 & 9.71 & 0.02 & 10.77 & 0.04 & 9.73 & 0.09 & 9.49 & 0.05 & 0.29 & 4.04\\
ZOA114948.692-640006.93 & J1149-64 & 1 & 296.241 & -1.934 & 0.57 & -63.4 & 42.66 & 11.73 & 0.02 & 10.42 & 0.02 & 9.73 & 0.02 & 11.08 & 0.29 & 9.84 & 0.16 & 9.13 & 0.16 & 2.44 & 4.38\\
ZOA141710.099-553238.77 & J1416-55 & 1$^*$ & 314.915 & 5.320 & 0.86 & 14.5 & 68.61 & 11.23 & 0.02 & 10.10 & 0.02 & 9.73 & 0.02 & 11.16 & 0.03 & 9.39 & 0.17 & 9.21 & 0.15 & 0.62 & 3.98\\
ZOA074141.201-223112.25 & J0741-22 & 1$^*$ & 238.558 & 0.239 & 0.80 & -20.1 & 48.26 & 11.31 & 0.02 & 10.24 & 0.02 & 9.74 & 0.02 & 11.03 & 0.07 & 9.30 & 0.25 & 9.68 & 0.02 & 0.66 & 3.74\\
ZOA161710.749-581844.59 & J1617-58 & 1$^*$ & 327.304 & -5.542 & 0.53 & 79.0 & 33.09 & 10.82 & 0.02 & 10.04 & 0.02 & 9.75 & 0.02 & 10.68 & 0.04 & 9.63 & 0.08 & 9.65 & 0.03 & 0.30 & 4.16\\
ZOA073008.083-220105.84 & J0730-22 & 1$^*$ & 236.817 & -1.851 & 0.77 & 22.5 & 104.61 & 11.51 & 0.02 & 10.04 & 0.02 & 9.78 & 0.02 & 10.71 & 0.51 & 9.25 & 0.11 & 9.20 & 0.07 & 1.81 & 3.69\\
ZOA075220.625-250840.47 & J0752-25A & 1$^*$ & 242.052 & 1.022 & 0.56 & 35.4 & 33.38 & 10.92 & 0.02 & 10.17 & 0.02 & 9.79 & 0.02 & 10.78 & 0.05 & 10.02 & 0.03 & 9.03 & 0.23 & 0.38 & 3.68\\
ZOA175453.601-342057.64 & J1755-34 & 1$^*$ & 356.354 & -4.464 & 0.37 & -15.7 & 33.99 & 11.04 & 0.02 & 10.35 & 0.02 & 9.83 & 0.02 & 10.77 & 0.16 & 9.73 & 0.19 & 9.28 & 0.16 & 0.70 & 4.86\\
ZOA182226.663-354035.70 & J1822-35 & 1$^*$ & 357.859 & -10.062 & 0.65 & 87.5 & 37.90 & 10.98 & 0.02 & 10.14 & 0.02 & 9.84 & 0.02 & 10.76 & 0.07 & 10.01 & 0.04 & 9.09 & 0.21 & 0.14 & 4.11\\
ZOA141604.868-651502.53 & J1416-65 & 1$^*$ & 311.644 & -3.821 & 0.48 & 69.7 & 36.60 & 11.23 & 0.02 & 10.32 & 0.02 & 9.90 & 0.02 & 10.48 & 0.26 & 10.10 & 0.04 & 9.38 & 0.14 & 0.77 & 4.31\\
ZOA105345.693-625013.17 & J1053-62 & 1$^*$ & 289.956 & -2.968 & 0.67 & -20.5 & 65.48 & 11.60 & 0.02 & 10.68 & 0.02 & 9.90 & 0.02 & 11.37 & 0.16 & 10.57 & 0.05 & 9.71 & 0.02 & 0.83 & 4.23\\
ZOA133732.784-585414.06 & J1337-58B & 2 & 308.867 & 3.436 & 0.28 & -16.5 & 33.07 & 11.33 & 0.02 & 10.31 & 0.02 & 9.92 & 0.02 & 10.93 & 0.13 & 9.93 & 0.06 & 9.70 & 0.05 & 1.09 & 4.30\\
ZOA191724.683+074909.02 & J1917+07 & 1$^*$ & 42.853 & -2.159 & 0.30 & 69.8 & 34.25 & 11.44 & 0.02 & 10.40 & 0.02 & 9.94 & 0.02 & 11.18 & 0.10 & 10.14 & 0.06 & 9.90 & 0.02 & 1.94 & 4.51\\
ZOA143927.759-552503.43 & J1439-55 & 1$^*$ & 317.910 & 4.281 & 0.22 & 44.9 & 30.55 & 11.17 & 0.02 & 10.41 & 0.02 & 9.97 & 0.02 & 10.83 & 0.11 & 9.68 & 0.21 & 9.10 & 0.26 & 0.64 & 4.11\\
\hline
\end{tabular}
   \end{small}
\end{table} 
\end{landscape}

\begin{figure}
 \centering
 \includegraphics[width=0.4\textwidth]{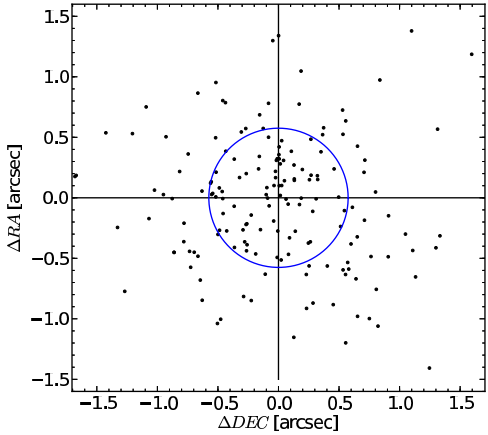}
 \caption[Comparison of IRSF and 2MASX positions]{Comparison of the  IRSF and 2MASX positions. The ellipse shows the $1\sigma$ dispersion in the positions.}
 \label{fig:comp2mass_pos}
\end{figure}

In Fig.~\ref{fig:comp2mass_pos} we show a comparison of the IRSF/SIRIUS and 2MASX positions. We find that the offset of $\Delta$RA of $-0\farcs02$ is small and consistent with zero, with a dispersion of $0\farcs61$, and $\Delta$DEC of $-0\farcs02$ with a dispersion of  $0\farcs54$. \cite{Riad2010} find similar values of $\Delta$RA $= -0\farcs01$ with a dispersion of $0\farcs42$, and $\Delta$DEC $=0\farcs11$ with a dispersion of $ 0\farcs39$. This is consistent with the expected combined error of 2MASX and IRSF/SIRIUS of $\sigma = 0\farcs44$ \citep{Riad2010}.  

\begin{figure*}
 \centering
 \includegraphics[width=0.48\textwidth]{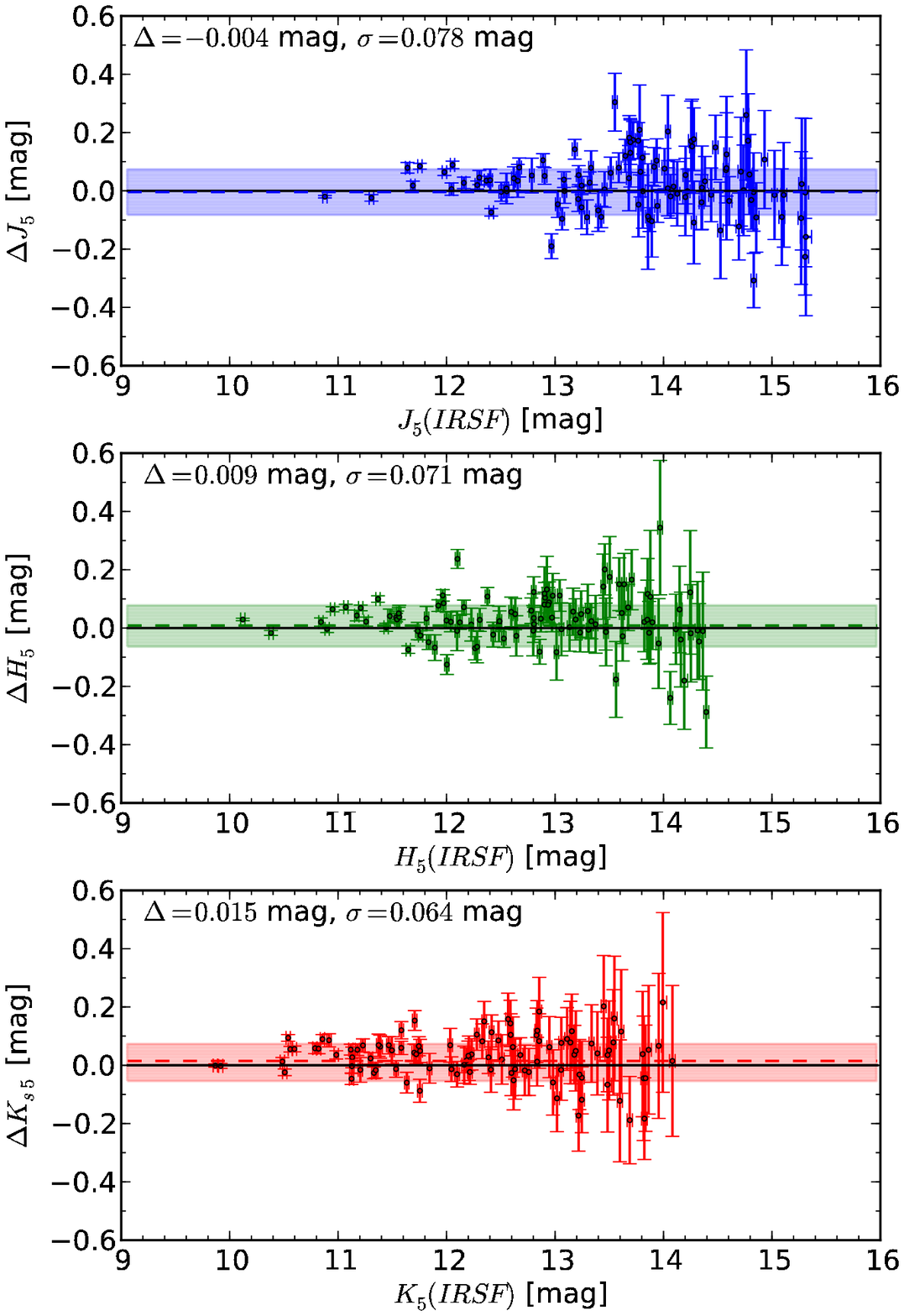}
 \includegraphics[width=0.48\textwidth]{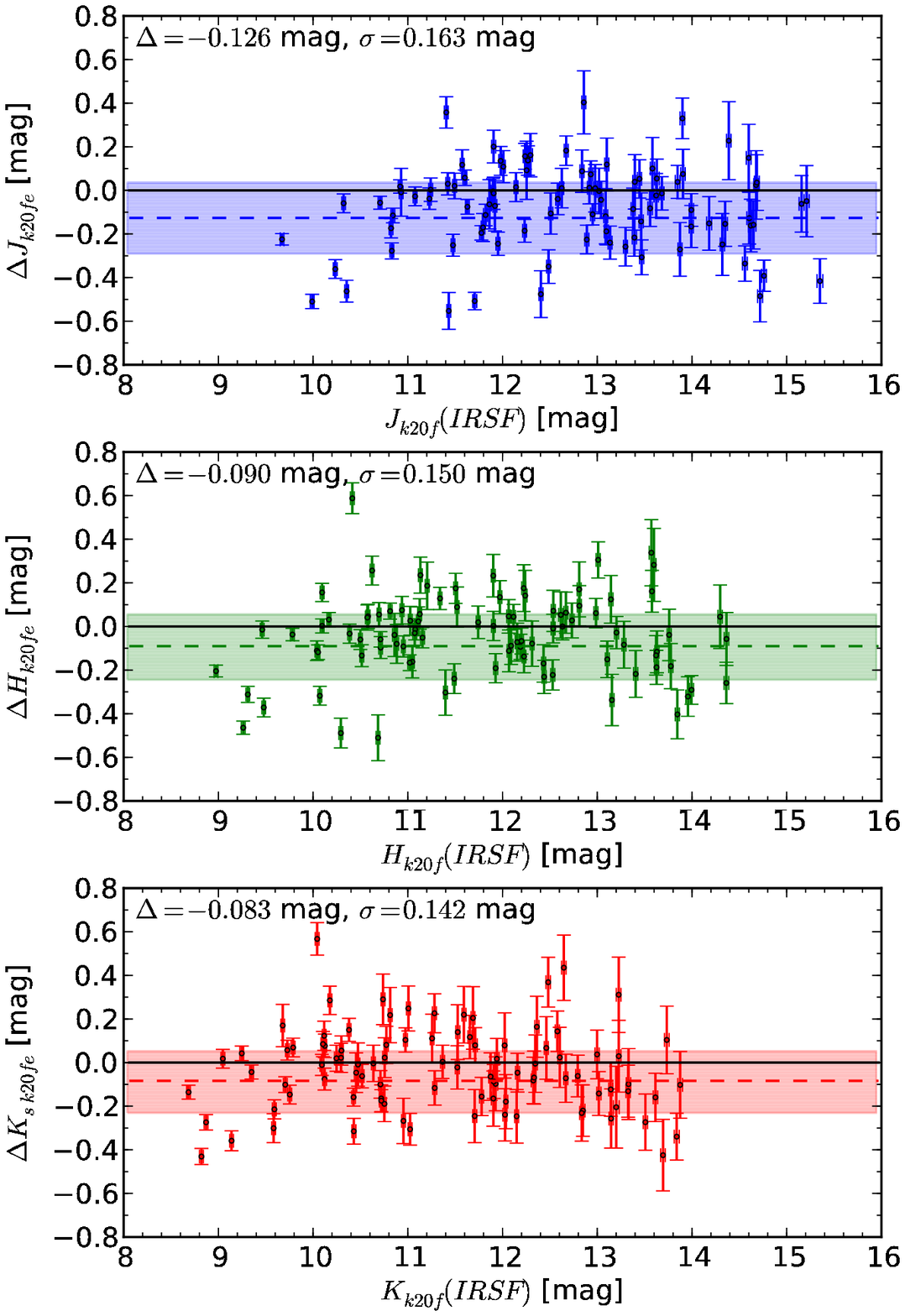}
 \caption[Comparison of IRSF and 2MASX aperture and isophotal magnitudes]{Comparison of the $5''$ circular aperture (\textit{left}) and  $K_s=20\m$ fiducial isophotal (\textit{right}) magnitudes in $J$ (\textit{top}), $H$ (\textit{middle}) and $K_s$ (\textit{bottom}) between the IRSF and 2MASX. The horizontal coloured strip shows the $1\sigma$ dispersion around the mean.  In both figures the horizontal coloured strip shows the $1\sigma$ dispersion around the mean (horizontal dashed line). The mean and standard deviation is given in the top left corner of each plot.}
 \label{fig:comp2mass}
\end{figure*}

Next, we compare the IRSF photometric parameters with the 2MASX values. The comparison is made for both the $5''$ aperture magnitudes and the $K_s=20$~mag~arcsec$^{-2}$ fiducial isophotal magnitudes in all three bands (see Fig.~\ref{fig:comp2mass}). The metric used is
\[
 \Delta m = m(2MASS) - m(IRSF),
\]
thus a positive quantity means that the 2MASS magnitude is fainter than the IRSF. All IRSF/SIRIUS magnitudes are transformed to the 2MASS standard via the transformation given by Nakajima \citetext{priv. comm; see Sect~\ref{sect:reduce}}. The mean offsets and dispersions are summarised in Table~\ref{tab:2masxcompare}; these are smaller for the $5''$ aperture than the isophotal magnitudes. This is  expected given the smaller apertures used, although they may be more affected by the different methods used for determining the position\footnote{2MASX uses the peak $J$ band pixel while our photometry is based on the centroid of the super co-add image.}. Note that seeing corrections are deemed negligible given the excellent observing conditions. 
\begin{table}
\begin{center}
 \caption[Comparison of 2MASX and IRSF/SIRIUS $5''$ and isophotal magnitudes]{Comparison of 2MASX and IRSF/SIRIUS $5''$ aperture and $K_s=20\m$ fiducial isophotal magnitudes.}
 \label{tab:2masxcompare}
 \begin{tabular}{ccccc}
    \hline
Filter & \multicolumn{2}{c}{$5''$ aperture} & \multicolumn{2}{c}{ $K_s=20\m$ fiducial} \\
   &  $\langle \Delta m \rangle$ & $\sigma$ &  $\langle \Delta m \rangle$ & $\sigma$ \\
   &  [mag] & [mag] &  [mag] & [mag] \\
  \hline
$J$  & -0.004 & 0.078 & -0.13 & 0.16 \\
$H$  & 0.009 & 0.071 & -0.09 & 0.15 \\
$K_s$& 0.015 & 0.064 & -0.08 & 0.14 \\
    \hline
 \end{tabular}
\end{center}
\end{table}

On average, we find that the IRSF isophotal magnitudes are fainter than the 2MASX magnitudes. This is not surprising: the higher resolution of IRSF/SIRIUS improves  the detection and subtraction of faint stars within the isophotal radius.
 The dispersion is $\sim2$ times that obtained by \cite{Riad2010} and \cite{Cluver+2008}, who find a dispersion of $0.095$~mag and $0.07$ respectively in $K_s$. \cite{Skelton2007} finds a higher dispersion of $0.19$, more consistent with our values. However, our larger dispersion may originate from the fact that our sample is composed mostly of spiral galaxies, many of which have extended faint disks, whereas Riad's sample contains many more elliptical galaxies. Moreover, our sample does cover a very wide range of Galactic longitude and latitude. Note also that the  $K_s=20\m$ fiducial isophotal magnitudes are measured in differently determined apertures.

\section{Discussion}
\label{sect:discuss}

\subsection{Detection Rates}
\label{sect:nircntparts}
The NIR catalogue contains all galaxies identified as possible \reply{NIR counterparts to the \HI~sources}, including some additional NIR galaxies where the counterpart was uncertain. Both the \HI~spectra and NIR images were used to aid in the counterpart identification. For instance a galaxy with a clear double horn profile was identified as an edge-on NIR galaxy and one with a flat-top profile was identified as a face-on galaxy. We also took the foreground extinction into account: e.g. if the \HI~profile was suggestive of a late-type spiral  but in a very high extinction region we allowed the galaxies to be more elliptical and bulgy in appearance because most of the low surface brightness disk will have been lost in the extinction.  In the cases of positive identification of the NIR counterpart, usually only the presence of a strong double horn profile and  clearly inclined galaxy allowed for confirmation of the counterpart. In the $578$ fields that were observed, a total of $557$ galaxies were identified in $421$ fields. $317$ Fields had a single counterpart, leaving $104$ fields with more than one NIR galaxy identified as a possible counterpart. Of these $104$ fields, a single counterpart could be ascertained for $46$ fields, while the NIR counterpart for $58$ fields remained ambiguous. Moreover, of the $317$ fields with only one possible counterpart, we consider $303$ to be confirmed based on visual inspection. The \HI\ sources with no identified NIR counterpart are listed in Table~\ref{tab:nir_nondetect}.

\begin{table}
 \centering
 \caption{H\,{\sevensize\bf I} Sources with no likely NIR counterpart}
 \label{tab:nir_nondetect}
\begin{center}
\begin{tabular}{lllll}
\hline
J0738-24 & J0741-22 & J0743-25A & J0746-35 & J0755-38 \\ 
J0758-36 & J0817-29A & J0821-45 & J0852-49 & J0902-40 \\ 
J0909-48B & J0912-53 & J0917-48 & J0927-49 & J0928-50 \\ 
J0931-43 & J0939-56 & J0945-48 & J0947-54 & J0949-49 \\ 
J0950-52 & J0951-53 & J0957-58 & J1000-58B & J1012-51 \\ 
J1014-54 & J1017-62 & J1024-61 & J1036-58 & J1049-55 \\ 
J1123-61 & J1134-59 & J1137-64 & J1143-57 & J1143-59 \\ 
J1149-62 & J1152-59 & J1206-61 & J1211-59 & J1216-65 \\ 
J1216-66 & J1221-61 & J1222-57 & J1230-59 & J1233-60 \\ 
J1233-65 & J1235-58 & J1242-57 & J1246-68 & J1249-58 \\ 
J1259-64 & J1304-58 & J1317-65 & J1329-60 & J1329-61 \\ 
J1341-64 & J1342-60 & J1343-65 & J1348-61 & J1358-59 \\ 
J1405-59 & J1412-56B & J1414-56 & J1414-62 & J1415-62 \\ 
J1416-55A & J1417-57 & J1419-58A & J1419-64 & J1422-57 \\ 
J1423-59 & J1424-60 & J1426-55 & J1426-60 & J1427-65A \\ 
J1429-58 & J1434-56B & J1438-60 & J1441-62 & J1442-57 \\ 
J1447-57 & J1452-55 & J1452-57 & J1501-59 & J1504-57 \\ 
J1511-60 & J1513-56 & J1522-52 & J1522-61 & J1528-52 \\ 
J1530-58 & J1531-52 & J1531-55 & J1534-56 & J1535-54 \\ 
J1536-56 & J1539-57 & J1539-59A & J1539-60 & J1541-52 \\ 
J1541-60 & J1542-55 & J1543-57 & J1545-56 & J1549-57 \\ 
J1551-59 & J1553-50 & J1558-53 & J1558-59 & J1600-56 \\ 
J1605-51 & J1605-59 & J1608-54 & J1612-52 & J1622-44A \\ 
J1622-53 & J1622-54 & J1629-42 & J1632-29A & J1638-49 \\ 
J1640-45 & J1643-49 & J1643-54 & J1644-42B & J1653-44 \\ 
J1656-44 & J1704-41 & J1705-29 & J1714-44 & J1719-37 \\ 
J1719-48 & J1721-37 & J1727-41 & J1727-49 & J1728-43 \\ 
J1739-24 & J1739-51 & J1749-32B & J1758-31 & J1758-33 \\ 
J1805-25 & J1807-06 & J1808-21 & J1814-35 & J1816-36 \\ 
J1817-04 & J1817-32 & J1820+07 & J1826-06 & J1834-30 \\ 
J1835-03 & J1844-11 & J1901+06 & J1907-00 & J1926+08 \\ 
J1930+11 & J1940+11 \\
\hline
\end{tabular}
\end{center}
\end{table}

\begin{figure}
 \begin{center}
 \includegraphics[width=0.45\textwidth]{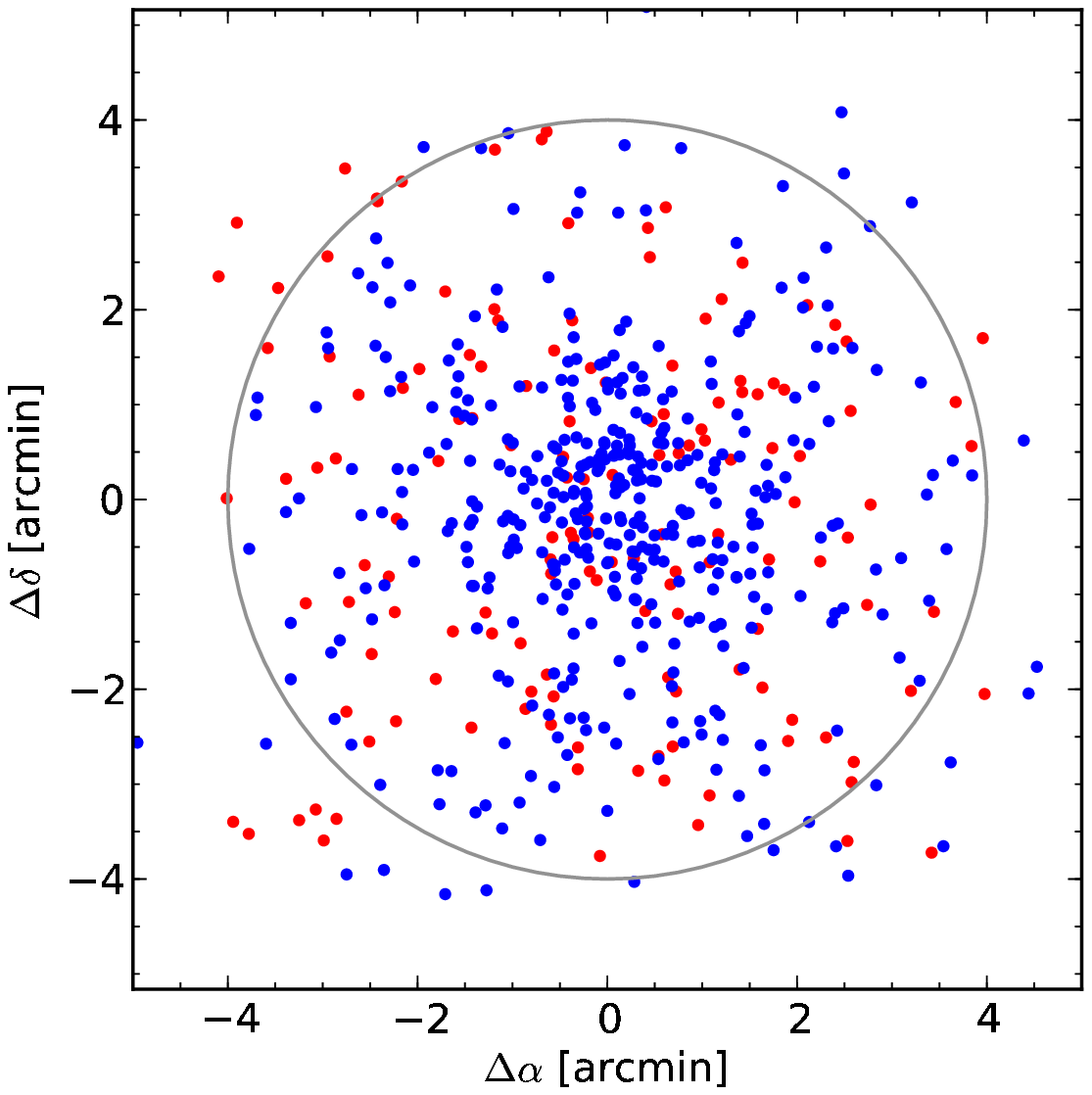}
 \caption[Positional offsets between \HI~positions and IRSF positions]{Positional offsets between \HI~positions and IRSF positions Confirmed counterparts are plotted in blue and unconfirmed counterparts in red. The large circle shows an offset of $4'$, which is the pixel size of the \HI~data. 94\% of the confirmed detections lie within $4'$, 77\% within $3'$.}
\label{fig:detect_pos_hi}
 \end{center}
 \end{figure}

The offsets between the \HI~and IRSF positions for the confirmed counterparts are plotted in Fig.~\ref{fig:detect_pos_hi}.  The large circle shows an offset of $4'$, which is the pixel size of the \HI~data. $98\%$ of the detections lie within $4'$ and $88\%$ are within $3'$. As a comparison, for the two parts of HIZOA Northern Extension (NE), \cite{Donley2005} confirm  $54-60\%$ 2MASS counterparts within $3'$.  In their final counterpart selection, \cite{Donley2005} exclude counterparts offset by $> 4'$ as statistically unlikely.  Our detection rate is slightly better than that of \cite{Donley2005} given that these NIR data are deeper than 2MASS and are better able to penetrate both the dust and stars of the Milky Way despite the fact that the NE fields lie at lower extinction and star density than much of the rest of the HIZOA fields. 

\begin{figure*}
 \begin{center}
 \includegraphics[width=\textwidth]{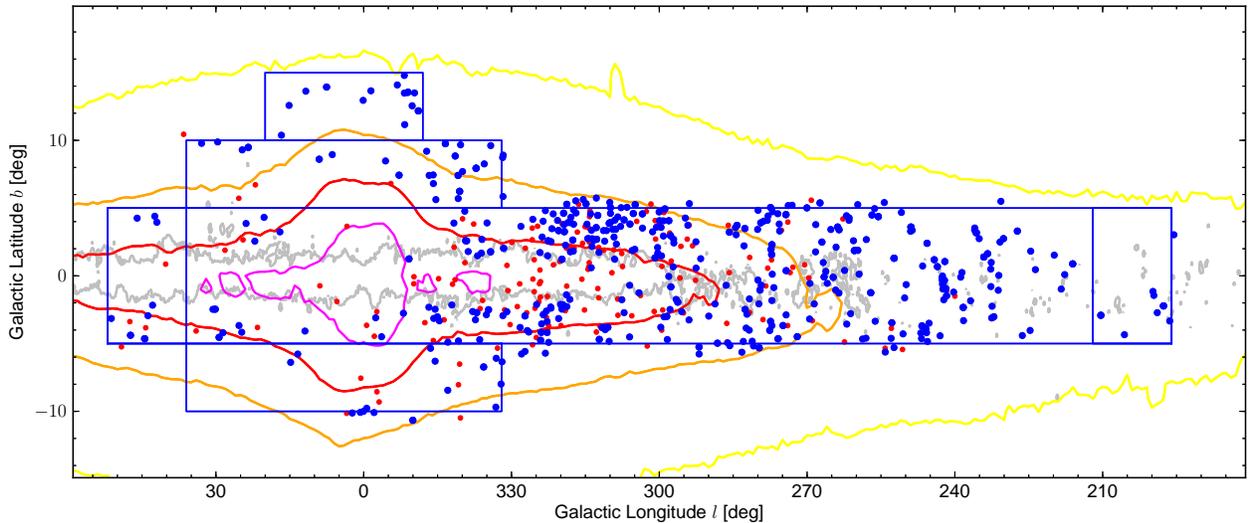}
 \caption[Spatial distribution of detected possible NIR counterparts]{Spatial distribution of $578$ detected and undetected possible NIR counterparts. The small red points show the $157$ fields for which no counterpart was identified. The large blue points indicate the $421$ fields in which one or more possible NIR counterparts for the \HI~source were identified. 2MASS PSC stellar density contours are overplotted with values of \reply{$\log (N_{({K_s} < 14 \m)}/\mbox{deg}^2) = 3.5, 4.0, 4.5$} and $5.0$  plotted in yellow, orange, red and magenta respectively. Very few detections are made at \reply{$\log (N_{({K_s} < 14 \m)}/\mbox{deg}^2) > 5.0$}. An extinction contour is plotted in grey for $A_{K_s} = 3.0$~mag.}
\label{fig:detect_distr}
 \end{center}
 \end{figure*}

Figure~\ref{fig:detect_distr} shows the spatial distribution of the $421$ galaxies with identified counterparts and the $157$ fields with no identifed counterpart. For  \reply{$\log (N_{({K_s} < 14 \m)}/\mbox{deg}^2) < 4.0$}, nearly all galaxies have NIR detections.  At stellar densities of \reply{$\log (N_{({K_s} < 14 \m)}/\mbox{deg}^2) > 4.5$}, the detection of galaxies correlates strongly with the extinction level. The likelihood of a detection is clearly determined by the foreground stellar density and for a given stellar density decreases at higher extinction. The amount of non-detections increases slightly at intermediate stellar densities. At slightly higher stellar densities, \reply{$4.5 < \log (N_{({K_s} < 14 \m)}/\mbox{deg}^2) < 5.0$}, most galaxies have NIR detections, except where the extinction level is very high ($A_{K_s} > 3.0$~mag). However, for very high star density, \reply{$\log (N_{({K_s} < 14 \m)}/\mbox{deg}^2) > 5.0$}, almost no counterparts are found, i.e. a ZoA of only $\pm 2\degr$ in latitude for the longitudes $\pm 60-70\degr$ around the Galactic Bulge.

\begin{figure}
 \begin{center} 
 \includegraphics[width=0.4\textwidth]{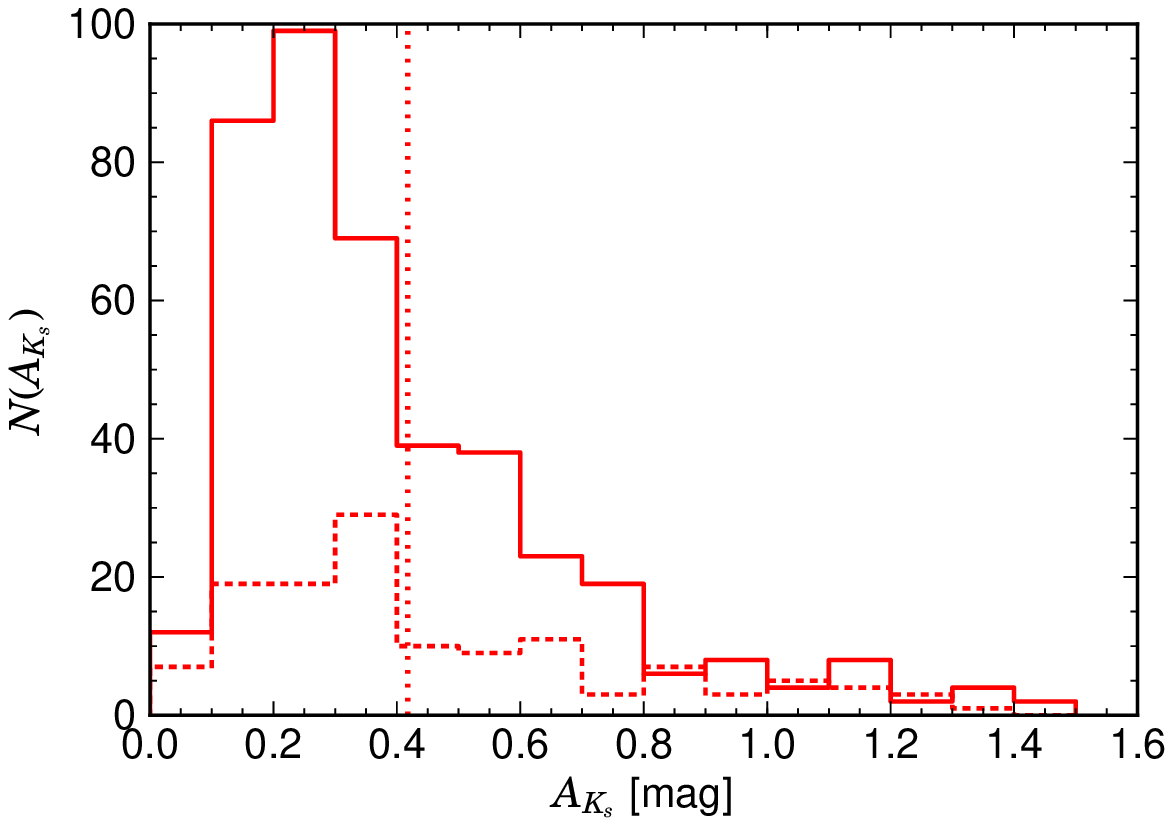}
 \includegraphics[width=0.4\textwidth]{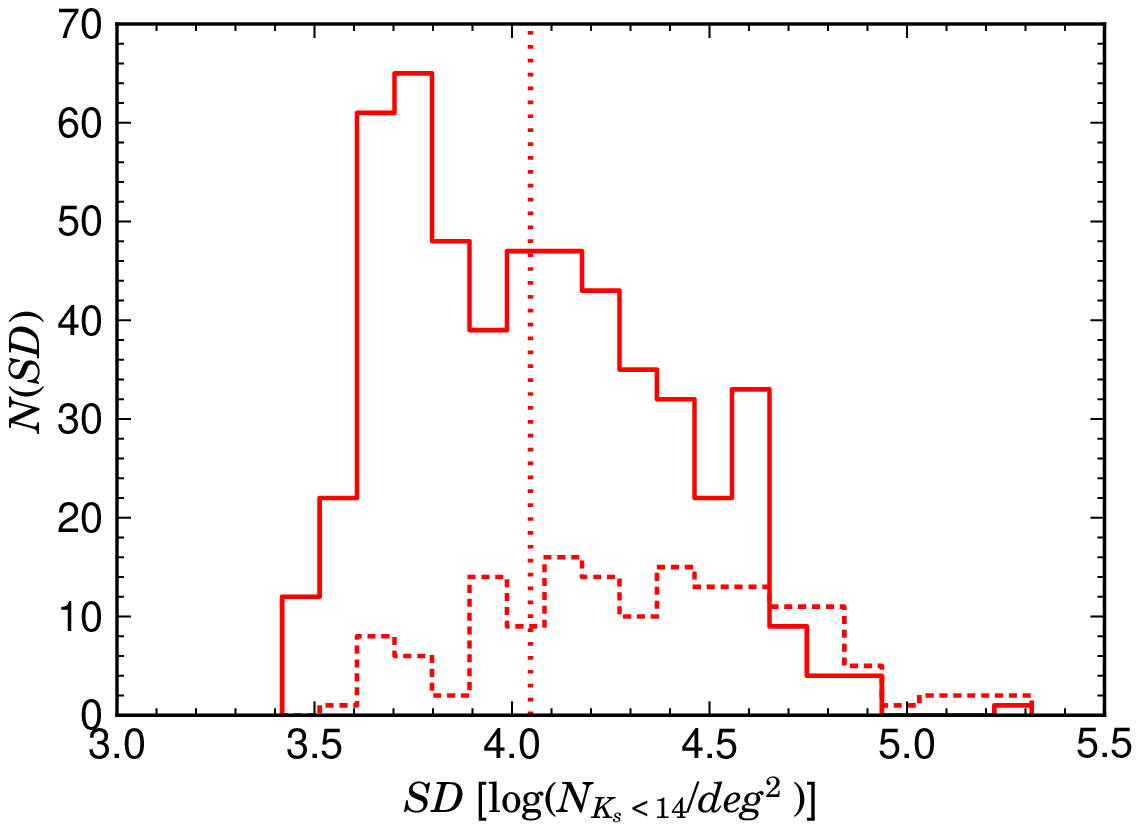}
 \caption[Extinction distribution]{\textit{Top} Number of fields with galaxies detected (solid line) and not detected (dotted line) as a function of $K_s$ band extinction. Only fields with $A_{K_s}<1.5$~mag are included; $0.90$~per~cent of the detected galaxies have $A_{K_s}>1.5$~mag. The vertical dotted line shows the mean extinction of $\langle 0.41\mbox{~mag} \rangle$. \textit{Bottom} As above but as a function of stellar density. The vertical dotted line shows the mean stellar density of $\langle 4.04\mbox{~mag} \rangle$.}
\label{fig:ext_dist}
 \end{center}
 \end{figure}

The  \cite{Schlegel1998} foreground extinction  of the detected galaxies, measured in the $K_s$ band, ranges between
$0.03$~mag and $4.48$~mag  with a mean extinction of $\langle 0.41\mbox{~mag} \rangle$.
Figure~\ref{fig:ext_dist} shows the number of galaxies detected and not detected as a function of
foreground extinction in the $K_s$ band. The distribution of detected and non-detected galaxies is also shown as a function of the stellar density in Fig.~\ref{fig:ext_dist}. The fraction of galaxies detected increases towards lower extinction and lower stellar density. However, two things are noteworthy: the overall distribution of non-detections is fairly flat in both plots, hence nearly independent of foreground extinction or star density.  These non-detections are most likely dwarfish blue low surface brightness galaxies that will not be picked up in any NIR survey, even far away from the Milky Way. \reply{Indeed, within the area with lower extinction and lower stellar density, the non-detections do on average have lower observed linewidths and \HI~masses than the \HI~sources with identified NIR counterparts. Moreover, the fraction of non-detections increases with decreasing \HI~linewidth and \HI~mass.} Secondly, the areal fraction of high Galactic extinction and/or high star density is small (see also Fig.10). 

\subsection{Magnitude and Colour Distribution}
\begin{figure}
 \begin{center}
 \includegraphics[width=0.5\textwidth]{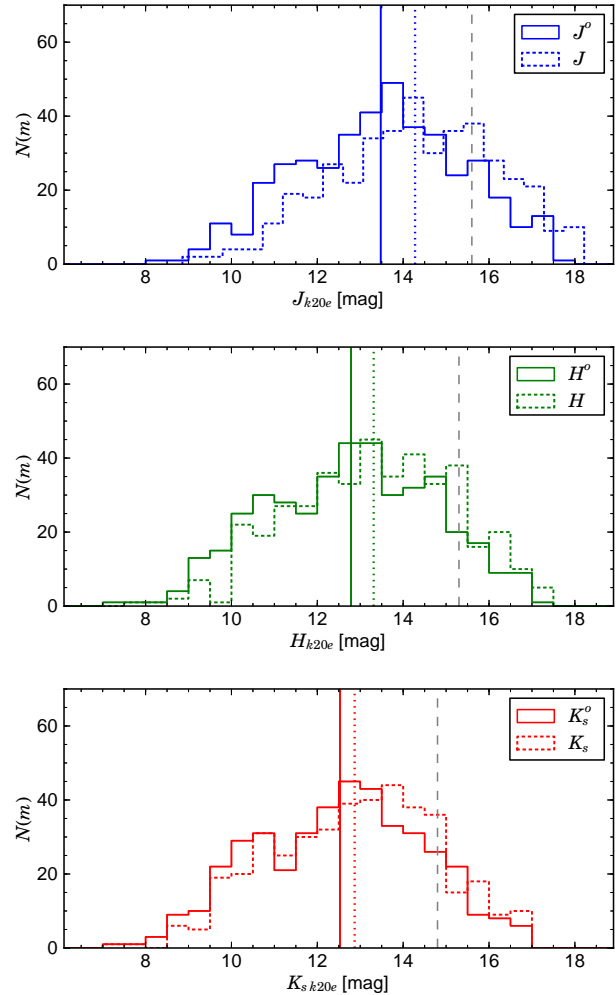}
 \caption[Magnitude distribution]{Distribution of $K_s=20$~mag~arcsec$^{-2}$~fiducial isophotal magnitudes in $J$ (\textit{top}, blue), $H$ (\textit{middle}, green) and $K_s$ (\textit{bottom}, red) in $0.5$~mag bins. The dotted histograms show the observed magnitudes and the filled solid histograms show the extinction-corrected magnitudes. The vertical dotted lines indicate the mean observed values and the vertical solid lines show the mean extinction-corrected magnitudes. The gray vertical dashed lines show the extinction-corrected completeness limits from \cite{Riad2010}.}
\label{fig:mag_dist}
 \end{center}
 \end{figure}

This section presents the magnitude and colour distributions of the $421$ detected galaxies. Figure~\ref{fig:mag_dist} shows the distribution of the observed (dotted histogram) and extinction-corrected (solid histogram) $K_s=20$~mag~arcsec$^{-2}$~fiducial isophotal magnitudes in $J$ (blue), $H$ (green) and $K_s$ (red). The magnitudes are corrected for foreground extinction based on the DIRBE/IRAS maps\footnote{we have not corrected the magnitudes for the changes in the isophotal shape due to extinction \citep[see]{Riad2010b}}, scaled by the factor $0.87$ derived in Sect.~\ref{Sect:nirext}. The mean observed and extinction-corrected magnitudes are listed in Table~\ref{tab:meanmag}; the mean extinction-corrected magnitudes are $0.80$, $0.52$ and $0.34$ magnitudes brighter than the mean observed magnitudes in $J$, $H$ and $K_s$ respectively. For the Norma Wall Survey (conducted on the IRSF using the same observational strategy and data reduction), \cite{Riad2010} determined magnitude completeness limits of $J=16.6$, $H=15.8$ and $K=15.4$~mag and extinction-corrected isophotal magnitudes of $J^o=15.6$, $H^o=15.3$ and $K_s^o=14.8$~mag in regions with dust obscuration below $A_{K_s} < 1.0$~mag and stellar densities less than \reply{$\log (N_{({K_s} < 14 \m)}/\mbox{deg}^2) = 4.71$}.

\begin{table}
\begin{center}
\caption[Mean observed and extinction-corrected magnitudes]{Mean observed and extinction-corrected $K_s=20$~mag\,arcsec$^{-2}$~fiducial isophotal magnitudes}
\label{tab:meanmag}
\begin{tabular}{ccc}
\hline
Filter & $\langle m \rangle $ & $\langle m^o \rangle$  \\
  &  [mag] & [mag] \\
\hline
$J$   & 14.28  &  13.48 \\
$H$   & 13.31  &  12.79 \\
$K_s$ & 12.87  &  12.53 \\ 
\hline
\end{tabular}
\end{center}
\end{table}

\begin{figure*}
 \begin{center}
 \includegraphics[width=\textwidth]{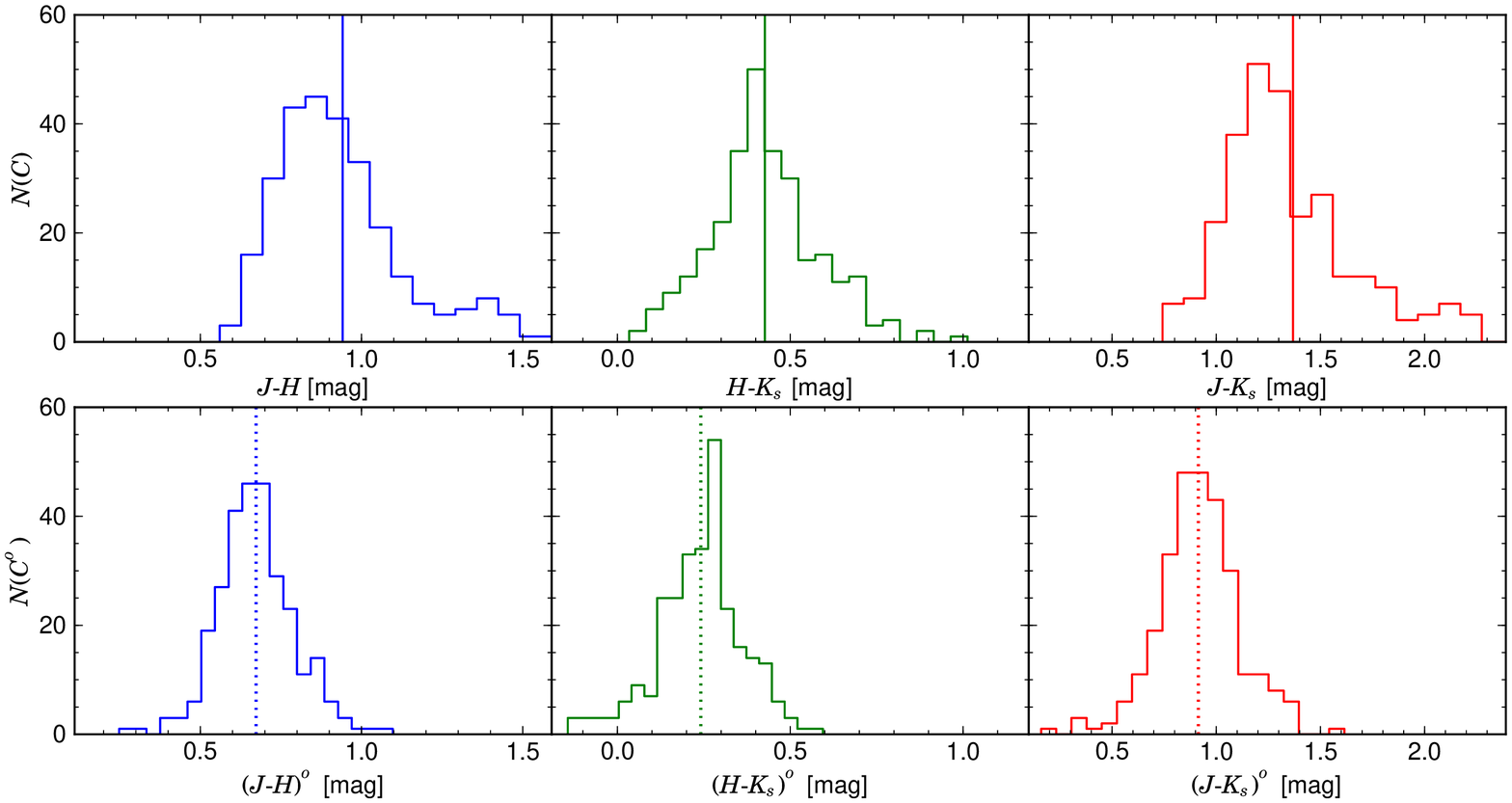}
 \caption[Colour distribution]{Histograms showing the distribution of NIR colours derived from the $5''$ aperture magnitudes. The \textit{top} panel shows the observed colour distributions which are skewed to the right (redder) due to selective extinction. The \textit{bottom} panel shows the extinction-corrected colour distributions which are more symmetric. $J-H$ is shown in blue, $H-K_s$ in green and $J-K_s$ in red. Vertical dotted lines indicate the mean values of each distribution.}
\label{fig:colour_dist}
 \end{center}
 \end{figure*}

The $5\arcsec$ aperture magnitudes are used to derive the NIR colours of the detected galaxies; the distribution of the observed colours is plotted in the \textit{top} panel of Fig.~\ref{fig:colour_dist} and that of the extinction-corrected colours in the \textit{bottom} panel. $J-H$ is plotted in blue, $H-K_s$ in green and $J-K_s$ in red. The mean values of each colour are plotted as vertical dotted lines. It is evident that the distributions of  observed colours are skewed towards the red end as a result of selective extinction. This is due to reddening by extinction. The extinction-corrected colour distributions are more symmetric and less dispersed. Table~\ref{tab:meancol} lists the mean values of the observed and extinction-corrected colours as well as the dispersion of each distribution. For comparison, the mean colours of the  2MASS Large Galaxy Atlas (LGA),  $\langle (J-H)^0 \rangle = 0.73$~mag, $\langle (H-K)^0 \rangle = 0.27$~mag and $\langle (J-K)^0 \rangle = 1.00$~mag, are also given \citep{Jarrett2000,Jarrett+2003}. The mean extinction-corrected colours of $\langle (J-H)^0 \rangle = 0.67$~mag, $\langle (H-K)^0 \rangle = 0.24$~mag and $\langle (J-K)^0 \rangle = 0.91$~mag are consistent with those of the LGA. \reply{We ignore the effects of morphological type on the distribution of the NIR colours and plot all the galaxies together. This is because we consider the majority of our galaxies to be of intermediate type (see Sect.~\ref{sect:types}), morphological types are harder to determine at low lattitudes and because NIR colours are known to vary only slightly with morphological type \cite[see][who plots the NIR colour distributions for the LGA for different morphological types]{Jarrett2000}.}  The dispersion of the extinction-corrected colours of $0.1-0.2$~mag is consistent with the dispersion in their distributions. The consistency between the mean values and dispersions of the extinction-corrected colours derived here and for the 2MASS LGA outside the ZoA adds credence to the new extinction correction derived in Sect.~\ref{Sect:nirext}.
\begin{table}
\begin{center}
\caption{Mean observed and extinction-corrected colours}
\label{tab:meancol}
\begin{tabular}{cccccc}
\hline
Colour & $\langle C \rangle $ & $\sigma_C $ & $\langle C^o \rangle$  & $ \sigma_{C^o}$ & $\langle C^o \rangle$ (LGA) \\
  &  [mag] & [mag] & [mag] & [mag] & [mag]\\
\hline
$J-H$   & 0.94  & 0.22  & 0.67 & 0.12 & 0.73 \\
$H-K_s$ & 0.43  & 0.16  & 0.24 & 0.12 & 0.27 \\
$J-K_s$ & 1.37  & 0.35  & 0.91 & 0.19 & 1.00 \\ 
\hline
\end{tabular}
\end{center}
\end{table}

\subsection{Usefulness for Tully-Fisher studies}
\reply{
For the 2MTF survey, \cite{Masters+2008} have already determined template relations for the Tully-Fisher relation in the 2MASS $J$, $H$ and $K_s$ bands which can be used for this data. We show here that the quality of our photometry is sufficient to be used for TF distances and peculiar velocities and so will be helpful in filling in the 2MTF ZoA. The dominant errors driving that of the TF peculiar velocities are  the absolute magnitude error, $\sigma_M$, the velocity width error, $|b|\sigma_W$ expressed in magnitudes where $b$ is the TF slope, and an intrinsic TF scatter term, $\epsilon_{int}$. Neglecting the covariances between the first two values arising from inclination corrections, the quadrature sum is approximately the total measurement error. We comment briefly on each of these terms.
}

\reply{
The absolute magnitude error is dominated by the measured apparent magnitude error, i.e. the errors in our photometry, as well as the error in the extinction correction. The measured errors in our magnitudes are typically $\sim 0.025$~mag, and for a brighter sample ($K_s \leq 11.25$~mag, the same selection for the 2MTF galaxies), the error is more typically $0.02$~mag. In comparison, the errors on the 2MASS galaxies are typically $0.1$~mag and $0.06$~mag for galaxies brighter than $K_s \leq 11.25$~mag. The $I$-band magnitude errors used in \cite{Springob+2007} are typically $<0.05$~mag.  For $A_{K_s}$ up to $1$~mag we expect the error in the extinction correction will be no larger than $0.04$~mag (based on the $3$~per~cent error on our extinction correction factor in Sect.~\ref{Sect:nirext}). We therefore conclude that the total magnitude error, under $0.1$~mag, is less than or, in the worst case, of the same order of the 2MASS magnitudes being used for the 2MTF. 
}

\reply{
Since the velocity width error is determined largely by the accuracy of the linewidths, we do not expand on it here, except to note that it will be affected by the accuracy of the inclination determined from our photometry. Our ellipticity (i.e. axis ratio) errors are typically $0.06$ which is better than the $0.1$ error estimated by \cite{Masters+2008} for the $J$ band axis ratio for 2MASS galaxies and comparable to the $0.04 - 0.06$ for the $I$ band axis ratios \citep{Springob+2007}.
}

\reply{
Finally, from \cite{Masters+2008}, the intrinsic scatter in the TF relation varies between $\sim 0.15$~mag and $1$~mag across $ 2 < \log W < 2.8$. 
}

 

\section{NIR Colours and Extinction}
\label{Sect:nirext}
The \cite{Schlegel1998} extinction maps derived from the DIRBE/IRAS data are not properly calibrated at low Galactic latitudes ($|b| < 5\dg$). Foreground contaminating sources are not removed from these maps at $|b| < 5\dg$ which can lead to an overestimate of the extinction. Moreover, \cite{Schlegel1998} assume an extinction law with $R_V \sim 3.1$, while the dust composition of giant molecular clouds within the Galactic disk is likely to  differ from less dense dust clouds further from the Galactic plane. Thus the extinction law and $R_V$ parameter should vary across the Galactic plane. Higher density regions consisting of larger grains typically have $R_V \sim 5$. For these regions, the difference in extinction as a function of wavelength is lower (large grains are grey absorbers) and would also imply an overestimation of the total extinction for a given colour. Nevertheless, the DIRBE/IRAS maps remain the best available means to estimate the extinction in the ZoA. Moreover, several studies have shown that while they do indeed overestimate the extinction, it is possible to correct for this. For example, \cite{Nagayama2004} measured the colour excess of giant stars in a region around PKS~1343-601 ($l,b = 309\fdg7, 1\fdg7$) and found the Schlegel values to be $33$~per~cent too high. 
\cite{Schroder+2007} looked at the  $I-J$, $J-K_s$ and $I-K_s$ NIR colours of galaxies in the same region and concluded that the extinction is $87$~per~cent of the DIRBE values, i.e. the DIRBE values overestimate the extinction by $15$~per~cent. Other independent studies have shown the true extinction to be $67-87$~per~cent of the DIRBE values \citep{Arce1999,Choloniewski2003,Dutra2003,Schroder2005,vanDriel2008}. Furthermore, in their study of Northern ZoA galaxies, \cite{vanDriel2008} find that the overestimate depends on the amount of extinction, increasing from a factor of $0.86$ for galaxies with $B$ band extinction in the range  $2 <A_B < 6$, to a factor of $0.69$ for the more heavily extincted galaxies with $6< A_B< 12$. The consistency of their values with those derived in the Southern ZoA suggests little or no variation with Galactic longitude.

\begin{figure}
 \centering
  \includegraphics[width=0.5\textwidth]{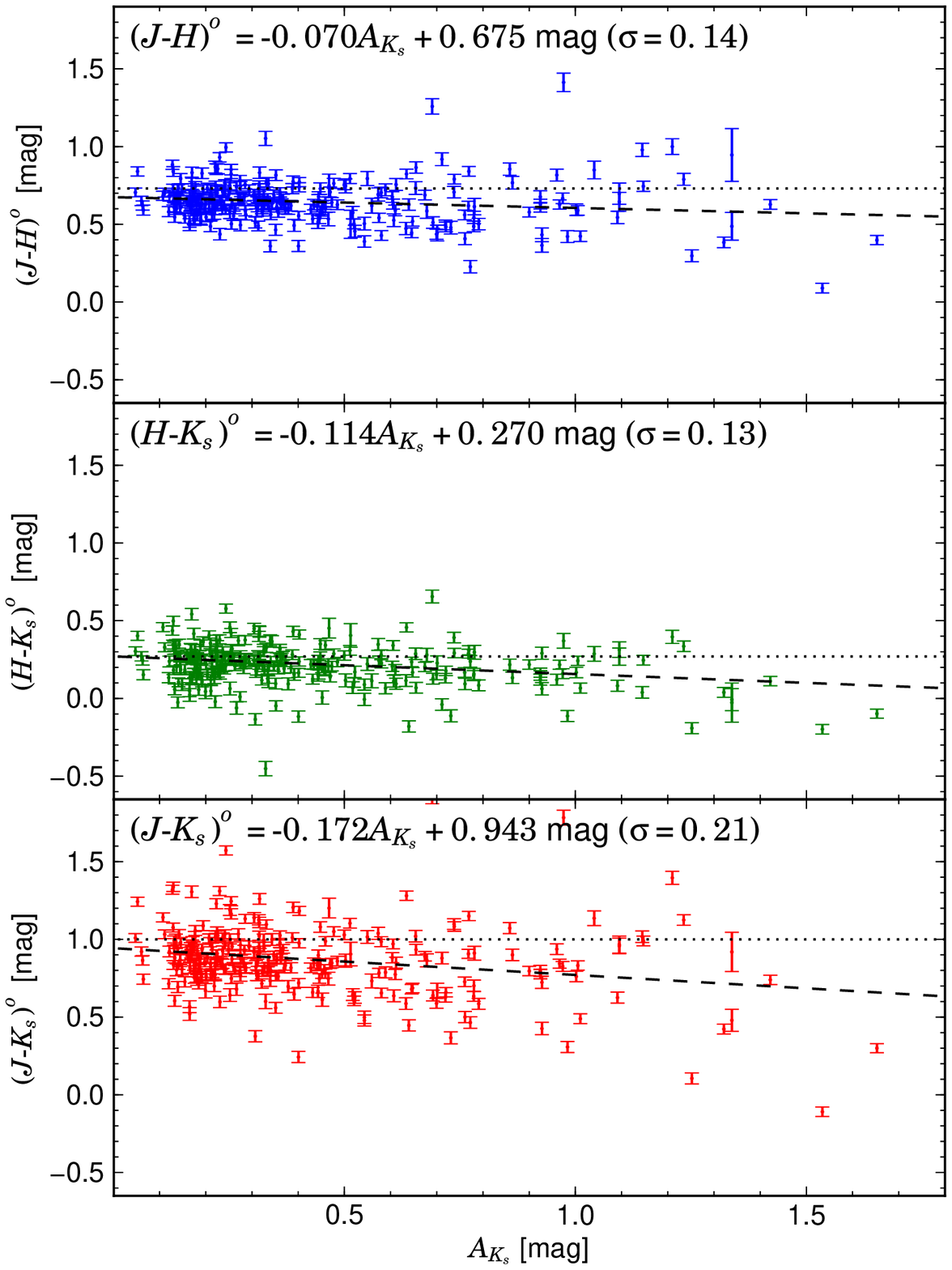}
 \caption[Extinction-corrected NIR colours as a function of $A_{K_s}$]{Extinction-corrected NIR colours plotted as a function of $K_s$ band extinction derived from the DIRBE/IRAS maps \citep{Schlegel1998}, from \textit{top} to \textit{bottom}: $(J-H)^o$ (blue), $(H-K_s)^o$ (green) and $(J-K_s)^o$ (red). The colours are corrected for extinction  based on the \cite{Schlegel1998} maps and \cite{Cardelli+1989} extinction laws. The dashed lines in each panel show a least squares linear fit to the colours and the fitted linear equation is given in the top left corner. The dotted lines show the mean intrinsic NIR galaxy colours determined by \cite{Jarrett+2003}.}
 \label{fig:NIR_colours}
\end{figure}

Since the NIR colours are nearly independent of galaxy type, the NIR colours of nearby galaxies provide an independent means to determine the extinction at low latitudes. We apply an extinction correction based on the DIRBE/IRAS values and look for residuals in the resulting colours; the true extinction corrected colours should be independent of extinction. Figure~\ref{fig:NIR_colours} shows the extinction-corrected NIR colours, $(J-H)^o$, $(H-K_s)^o$ and $(J-K_s)^o$, as a function of the $K_s$ band extinction, $A_{K_s}$, for all NIR detections with reliable photometry in all three bands. The galaxies cover  a wide range of Galactic latitude and longitude.  The colours were determined from the $5''$ aperture magnitudes. The DIRBE/IRAS maps clearly overestimate the extinction at these low latitudes. The dashed lines show a linear least squares fit to the data. There is a clear trend for the colours of galaxies at high extinction to be bluer than expected. The fits give:
\[
\label{eq:colours}
( J - H  )^o  =  (-0.070 \pm 0.027) A_{K_s} + (0.675 \pm 0.013), \]
\[
( H - K_s)^o  =  (-0.114 \pm 0.026) A_{K_s} + (0.270 \pm 0.013), \mbox{~and} \]
\[
( J - K_s)^o  =  (-0.172 \pm 0.044) A_{K_s} + (0.943 \pm 0.021),
\]
with an rms of $0.14$ mag, $0.13$~mag and $0.21$~mag respectively.

Following the method first derived by \cite{Schroder+2007}, we assume that the Schlegel values, $A_{K_s}$,  overestimate (underestimate) the true extinction, $ \widetilde{A_{K_s}}$, by a constant multiplicative factor $f$, i.e.
\[
 \widetilde{A_{K_s}} = f A_{K_s}.
\]
The  correction factor is related to the slope of the linear fit by
\[
 f = a \left( \frac{E(C)}{A_{K_s}} \right)^{-1} + 1,
\]
where $E(C)/{A_{K_s}}$ is the ratio of selective to total extinction which is obtained by integrating the extinction law of \cite*{Cardelli+1989} over each of the SIRIUS $J$, $H$ and $K_s$ bandpasses:
\[
A_J= 0.863 E(B-V),
\]
\[
A_H= 0.570 E(B-V ),
\]
\[
A_{K_s}= 0.368 E(B-V ).
\]
From the linear fits we derive
\[
f_{(J-H)}   =  0.913 \pm 0.030, \]
\[
f_{(H-K_s)} =  0.793 \pm 0.042,\mbox{~and} \]
\[
f_{(J-K_s)} =  0.784 \pm 0.046,
\]
however, only two are independent. We choose to take the weighted mean of $f_{(J-H)}$ and $f_{(H-K_s)}$ as these colours are least affected by $k$-corrections \citep[see e.g.][]{Poggianti1997} and have the smallest errors. 
Thus we get
\[
 f = 0.87 \pm 0.03.
\]
i.e. the true extinction is 13~per~cent lower than predicted by the DIRBE/IRAS values, averaging across the entire southern ZoA. This is consistent with the value of \cite{Schroder+2007} in the region of PKS 1343-601 ($l,b = 309.7\degr, 1.7\degr$).

\subsection{Discussion of Possible Sample Biases}
We investigate whether the observed dependence of colour on the extinction is a result of a selection effect or bias in our sample. We consider three effects:
\begin{enumerate}
 \item The loss of intrinsically red galaxies at higher extinction due to selective absorption near the completeness limit of the NIR observations. Adopting the magnitude limits of \cite{Riad2010}: $16.6$, $15.8$ and $15.4$~mag in $J$, $H$ and $K_s$ respectively, we calculate that a limit of $14.83$~mag in the $K_s$ band will be sensitive to $J-K$ colours within $0.5$~mag of the mean value even with an extinction of $A_{K_s}$ of $1.4$~mag. Similarly a cut-off of $15.03$~mag in $K_s$ applies for the $H-K$ colours. Finally, for $J-H$, a cut-off of $H = 15.27$ allows us to detect galaxies with colours up to $0.5$~mag from the mean with an extinction of $A_{K_s} = 1.5$~mag. After selecting sub-samples based on cut-offs of $K_s = 14.75$~mag for $J-K_s$ and $H-K_s$ and $H = 15.0$~mag for $J-H$, we find no significant changes in the observed relation between the Schlegel extinction-corrected colours and the extinction. Therefore, we are confident that this is not a result of the completeness of the sample.
 \item The effect of redshift on NIR colours, i.e. the $k$ correction. \cite{Poggianti1997} give small linear corrections with $z$ which amount to $<0.03$~mag for our galaxies. Limiting the sample to only those galaxies with confirmed counterparts so their velocities are known and selecting only those with $4\,000 < v < 6\,000$~\kph, we purposefully select a sample that will exaggerate this effect. However, we derive similar correction values for this sub-sample, indicating that it is not the major contributor to the observed gradients in the colour-extinction relations.
 \item Small variations in the intrinsic colours of the detected galaxies as a function of galaxy type. The $J-H$ colour drops from the nominal value of $1.0$~mag to $0.7$~mag towards later type galaxies (Sdm). To investigate this effect, we select from the \HI~confirmed sample only those galaxies with \reply{inclination-corrected} linewidths $\log W < 2.2$, which are likely to be later type Sdm galaxies. Again we find no significant difference in the derived correction factor. \reply{We further test this by using a sample of galaxies that are classed as later type based on visual inspection of the NIR images (T-types 4 and 5). There is no significant difference in the correction factor for this sample.}
\end{enumerate}

\subsection{Variation with Galactic Latitude and Longitude}
\label{sect:latvary}
Because this sample covers such a large area along the Southern Galactic plane, we can use it to investigate the extinction in different regions of the Galactic plane. In the following section we determine the correction factor in several latitude and longitude bins.

\begin{figure}
 \centering
  \includegraphics[width=0.45\textwidth]{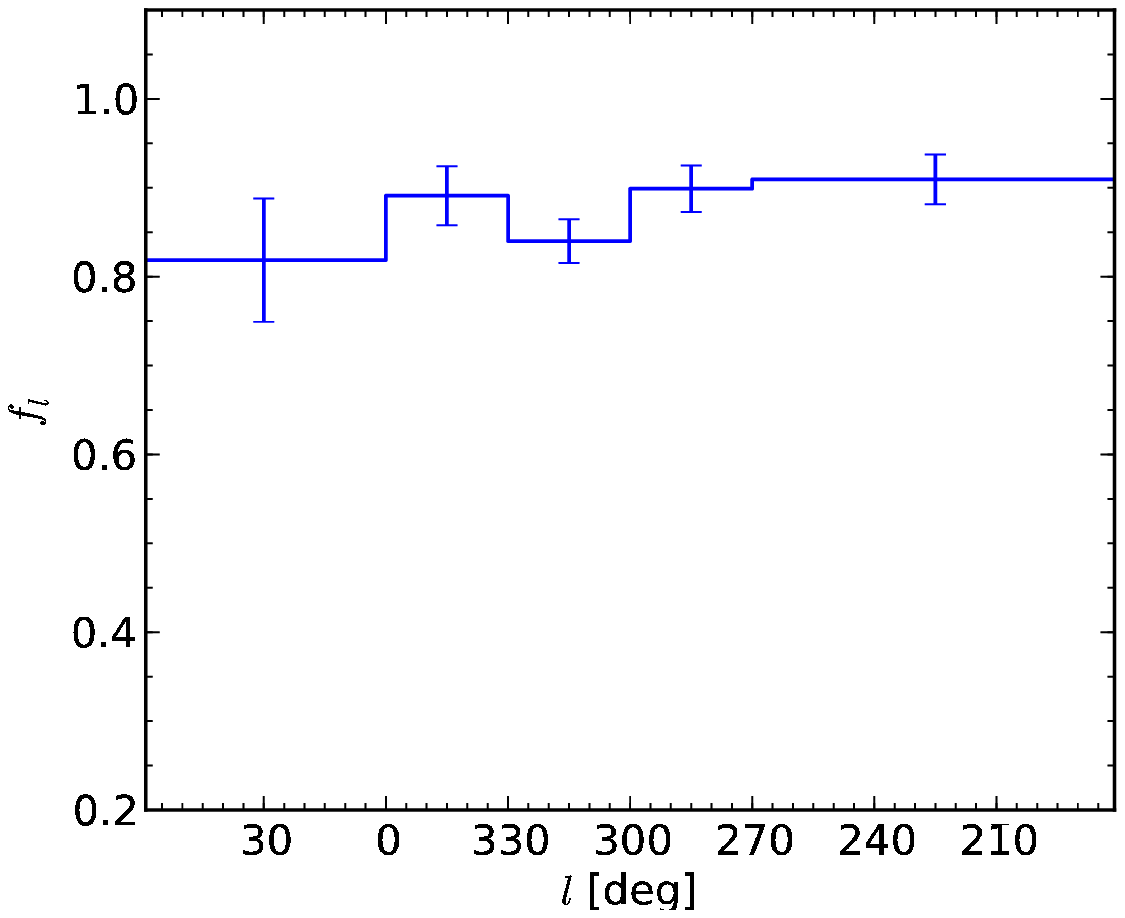}
  \includegraphics[width=0.45\textwidth]{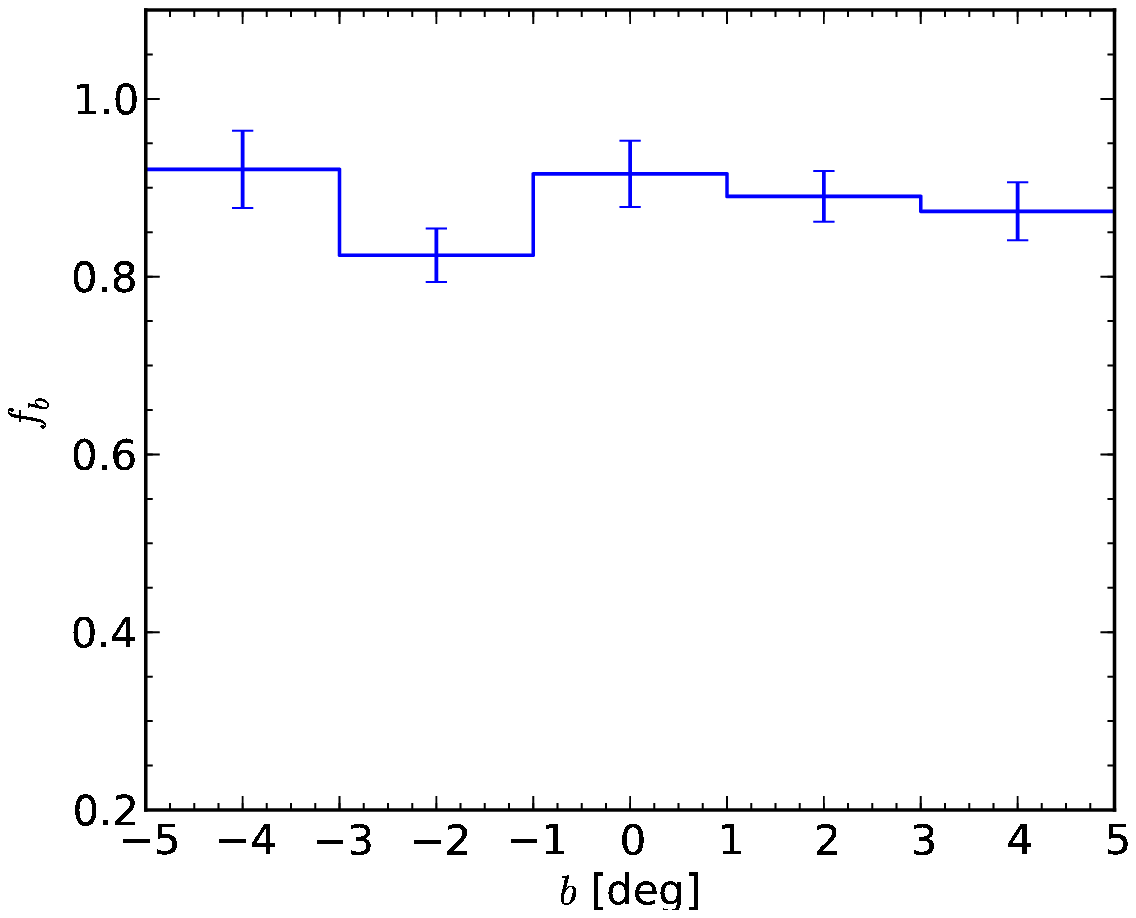}
 \caption{DIRBE/IRAS extinction correction factor as a function of Galactic longitude, $l$ (\textit{Top})and Galactic latitude,  $b$ (\textit{Bottom}). }
 \label{fig:f_l}
\end{figure}

Fig.~\ref{fig:f_l} shows the extinction factor derived for several Galactic longitude and latitude bins. The size of the bins is restricted by the number of galaxies we have. Each bin contains $15-60$ galaxies. For this case, we fit only the slope and keep the intercept fixed at the LGA values for the extinction-corrected NIR colours. We see no significant variation in the correction factor as a function of longitude or latitude. There is some variation within the errors. This is consistent with the results of \cite{vanDriel2008}, which suggest continuity of the overestimation across Galactic longitude. Initial work by Burstein and Heiles \citep[see e.g.][]{Heiles1976,Burstein1978,Burstein1982} suggested a variation with longitude based on the observed changes in the relationships between the \HI~gas-to-dust ratio and galaxy counts towards the Galactic bulge.

 \reply{It is interesting to note that this 13 per cent lower DIRBE/IRAS extinction value nearly matches the value of 13 found by \cite{Schlafly.et.al.2010} in a completely independent  analysis of the blue tip of main sequence turn-off stars, and that of 14 per cent found by \cite{Schlafly.Finkbeiner.2011} in a different analysis  of the colours of stars with spectra in the Sloan Digital Sky Survey. Both studies are based on colours determined from other filterbands and in regions far away from the Galactic Plane. Their comparable correction factors may well suggest that we have found the same effect, and this is in fact not induced by different dust properties at lowest Galactic latitudes.}

\section{Summary}
\label{sect:summ}
Deep NIR follow-up observations of HIZOA galaxies within $6000$~\kph~are
presented. $J$, $H$ and $K_s$ images of $578$ targets were obtained using the
$1.4$\,m IRSF telescope with the SIRIUS camera. The combined three-colour NIR
fields were visually searched for counterparts of the \HI~galaxies and following
the identification of possible counterpart galaxies, careful subtraction of
foreground stars was done. Surface photometry was performed on the
star-subtracted images to produce a photometric catalogue of $555$ NIR galaxies.
A range of photometric parameters are determined for each source and used to
compile the final NIR catalogue; these include the ellipticity, position angle,
isophotal magnitudes and extrapolated total magnitudes in all three NIR bands. 

A comparison with 2MASX positions found no significant offset, with dispersions of the order of $ 0\farcs6$ and $ 0\farcs4$
for RA and Dec respectively. The magnitudes at small apertures all agreed very well with dispersions of  0.07 mag. There was a slightly larger offset for $K_s=20$~mag~arcsec$^{-2}$ fiducial isophotal magnitudes, which were fainter as derived from the IRSF photometry compared to 2MASX ($-0.13$, $-0.09$ and $-0.08$~mag for the three NIR bands). This conforms  well with the expectations because of the improved star subtraction that is possible with the higher resolved images. The dispersion hovers around 0.15 mag when compared to 2MASX. This again is understood in that the deeper IRSF images will often allow a better recovery of extended low-surface brightness disks of this spiral galaxy sample. 

\reply{
We have shown that the quality of the photometry is sufficient for use with the TF relation to determine distances and peculiar velocities. The expected magnitude errors are less than $0.1$~mag and may be $<0.05$~mag in regions of lower extinction. The errors on the axis ratios are typically $0.06$ which will not be prohibitive in determining inclinations for velocity width corrections.
}

The detection of NIR counterparts has proved quite successful. For \reply{$\log (N_{({K_s} < 14 \m)}/\mbox{deg}^2) < 4.0$}, nearly all galaxies have NIR detections, though almost none are found for very high star density, \reply{$\log (N_{({K_s} < 14 \m)}/\mbox{deg}^2) > 5.0$},  i.e. resulting in a ZoA of only $\pm 2\degr$ in latitude for the longitudes $\pm 60-70\degr$ around the Galactic Bulge.  Within that higher stellar density range, the detection of galaxies correlates strongly with the extinction level. As shown with Fig.~10,  most galaxies still have a NIR counterpart for  \reply{$4.5 < \log (N_{({K_s} < 14 \m)}/\mbox{deg}^2) < 5.0$}. However, this does not hold when  the extinction level is very high ($A_{K_s} > 3.0$~mag). Overall, the deeper NIR follow-up imaging project of HIZOA galaxies observations shows a vast improvement compared to 2MASX, which recovered only 130 certain 2MASX counterparts for the full HIZOA catalogue of $\sim 1100$ galaxies.

We have used the NIR colours of all the detected galaxies to investigate the
extinction in the Galactic plane, to test whether the dust properties change at these high dust column densities.  The ratio of the true extinction to the
DIRBE/IRAS extinction \citep{Schlegel1998}  was found to be $0.87$ across the whole 
HIZOA survey region.  This value showed no significant variation with Galactic longitude and latitude. This value is completely compatible with the values found by \cite{Schlafly.Finkbeiner.2011} which were obtained based on an analysis of stellar data. The result here might be an independent confirmation, rather than being due to a difference in dust properties at low Galactic latitudes.

\section*{Acknowledgments}

\textit{We acknowledge the HIZOA survey team for early access to the data. We also thank the numerous observers that have helped with completing this survey and T. Jarrett and A. Schr\"{o}der for insightful discussions on the surface photometry. We thank the anonymous referee for helpful suggestions. The work of WLW was based on research generously supported by the South African SKA project. WLW, PAW and RKK acknowledge financial support from UCT and the NRF. This work is based on observations obtained at the South African Astronomical Observatory.}


\bibliographystyle{mn2e}
\bibliography{refs.bib}

\bsp


\label{lastpage}

\end{document}